\documentclass[aps, pre, floatfix, twocolumn]{revtex4}
\usepackage{amsmath}
\usepackage{verbatim}
\usepackage[english]{babel}
\usepackage[T1]{fontenc}
\usepackage{graphicx}
\usepackage{listings}
\usepackage{lscape}
\usepackage{makeidx}
\usepackage{url}
\usepackage{subfigure}
\usepackage{array}
\usepackage{amsfonts}
\usepackage{enumerate}
\usepackage{MnSymbol}
\usepackage{psfrag}

\newcommand{\lpar}{\left(}
\newcommand{\rpar}{\right)}
\newcommand{\lbk}{\left \lbrack}
\newcommand{\rbk}{\right \rbrack}
\newcommand{\lbc}{\left \lbrace}
\newcommand{\rbc}{\right \rbrace}
\newcommand{\ga}{{\alpha}}

\newcommand{\beq}[1]{\begin{equation}\label{#1}}
\newcommand{\eeq}{\end{equation}}
\newcommand{\bal}[1]{\begin{align}\label{#1}}
\newcommand{\eal}{\end{align}}
\newcommand{\eref}[1]{Eq.~\eqref{#1}}
\newcommand{\figref}[1]{\ref{#1}}
\newcommand{\sref}[1]{(\ref{#1})}
\newcommand{\vecc}[1]{\mathbf{#1}}

\psfrag{ux}[Bc][Bc]{\scriptsize\hspace{-0.0\textwidth} $u_{x,0}$}
\psfrag{dfdf}[Bc][Bc]{\scriptsize\hspace{-0.0\textwidth} $\langle \delta f_i \delta f_i \rangle$}
\psfrag{df0dfi}[Bc][Bc]{\scriptsize\hspace{-0.0\textwidth} $\langle \delta f_0 \delta f_i \rangle$}
\psfrag{dmdm}[Bc][Bc]{\scriptsize\hspace{-0.0\textwidth} $\langle \delta M^a \delta M^a \rangle/\rho$}
\psfrag{dm0dma}[Bc][Bc]{\scriptsize\hspace{-0.0\textwidth} $\langle \delta M^0 \delta M^a \rangle/\rho$}

\psfrag{XXXXXXXXXXf0f0}[Bl][Br]{\scriptsize\hspace{-0.075\textwidth} $\langle \delta f_0 \delta f_0 \rangle$}
\psfrag{XXXw0}[Bl][Br]{\scriptsize\hspace{-0.075\textwidth} $w_0 = \frac{4}{9}$}
\psfrag{XXXf0}[Bl][Br]{\scriptsize\hspace{-0.075\textwidth} $f_0^0$}

\psfrag{XXXXXXXXXXf1f1}[Bl][Br]{\scriptsize\hspace{-0.075\textwidth} $\langle \delta f_1 \delta f_1 \rangle$}
\psfrag{XXXf2f2f4f4XXX}[Bl][Br]{\scriptsize\hspace{-0.075\textwidth} $\langle \delta f_2 \delta f_2 \rangle$}
\psfrag{XXXf3f3}[Bl][Br]{\scriptsize\hspace{-0.075\textwidth} $\langle \delta f_3 \delta f_3 \rangle$}
\psfrag{XXXw1}[Bl][Br]{\scriptsize\hspace{-0.075\textwidth} $w_{1...4} = \frac{1}{9}$}
\psfrag{XXXf1}[Bl][Br]{\scriptsize\hspace{-0.075\textwidth} $f_1^0$}
\psfrag{XXXf2f4XXX}[Bl][Br]{\scriptsize\hspace{-0.075\textwidth} $f_2^0 = f_4^0$}
\psfrag{XXXf3}[Bl][Br]{\scriptsize\hspace{-0.075\textwidth} $f_3^0$}

\psfrag{XXXf2}[Bl][Br]{\scriptsize\hspace{-0.075\textwidth} $f_2^0$}

\psfrag{XXXf5f5f8f8XXX}[Bl][Br]{\scriptsize\hspace{-0.075\textwidth} $\langle \delta f_5 \delta f_5 \rangle$}
\psfrag{XXXf6f6f7f7XXX}[Bl][Br]{\scriptsize\hspace{-0.075\textwidth} $\langle \delta f_6 \delta f_6 \rangle$}
\psfrag{XXXw5}[Bl][Br]{\scriptsize\hspace{-0.075\textwidth} $w_{5...8} = \frac{1}{36}$}
\psfrag{XXXf5f8XXX}[Bl][Br]{\scriptsize\hspace{-0.075\textwidth} $f_5^0 = f_8^0$}
\psfrag{XXXf6f7XXX}[Bl][Br]{\scriptsize\hspace{-0.075\textwidth} $f_6^0 = f_7^0$}
\psfrag{XXXXXXXXf0f1}[Bl][Br]{\scriptsize\hspace{-0.065\textwidth} $\langle \delta f_0 \delta f_1 \rangle$}
\psfrag{XXXf0f2}[Bl][Br]{\scriptsize\hspace{-0.065\textwidth} $\langle \delta f_0 \delta f_2 \rangle$}
\psfrag{XXXf0f3}[Bl][Br]{\scriptsize\hspace{-0.065\textwidth} $\langle \delta f_0 \delta f_3 \rangle$}
\psfrag{XXXf0f4}[Bl][Br]{\scriptsize\hspace{-0.065\textwidth} $\langle \delta f_0 \delta f_4 \rangle$}
\psfrag{XXXf0f5}[Bl][Br]{\scriptsize\hspace{-0.065\textwidth} $\langle \delta f_0 \delta f_5 \rangle$}
\psfrag{XXXf0f6}[Bl][Br]{\scriptsize\hspace{-0.065\textwidth} $\langle \delta f_0 \delta f_6 \rangle$}
\psfrag{XXXf0f7}[Bl][Br]{\scriptsize\hspace{-0.065\textwidth} $\langle \delta f_0 \delta f_7 \rangle$}
\psfrag{XXXf0f8}[Bl][Br]{\scriptsize\hspace{-0.065\textwidth} $\langle \delta f_0 \delta f_8 \rangle$}

\psfrag{XXXXXXXXXXXXXXXXXRR}[Bl][Br]{\scriptsize\hspace{-0.1\textwidth} $\langle \delta \rho \delta\rho \rangle$}
\psfrag{XXXJXJX}[Bl][Br]{\scriptsize\hspace{-0.10\textwidth} $\langle \delta j_x \delta j_x \rangle$}
\psfrag{XXXJYJY}[Bl][Br]{\scriptsize\hspace{-0.10\textwidth} $\langle \delta j_y \delta j_y \rangle$}
\psfrag{XXXPXXMPYY}[Bl][Br]{\scriptsize\hspace{-0.1\textwidth} $\langle \lpar \delta \Pi_{xx-yy} \rpar^2 \rangle$}
\psfrag{XXXPXY}[Bl][Br]{\scriptsize\hspace{-0.10\textwidth} $\langle \lpar \delta \Pi_{xy} \rpar^2 \rangle$}
\psfrag{XXXPXXPPYY}[Bl][Br]{\scriptsize\hspace{-0.10\textwidth} $\langle \lpar \delta \Pi_{xx+yy} \rpar^2 \rangle$}
\psfrag{XXXG1G1}[Bl][Br]{\scriptsize\hspace{-0.10\textwidth} $\langle \delta q_x \delta q_x \rangle$}
\psfrag{XXXG2G2}[Bl][Br]{\scriptsize\hspace{-0.10\textwidth} $\langle \delta q_y \delta q_y \rangle$}
\psfrag{XXXG3G3}[Bl][Br]{\scriptsize\hspace{-0.10\textwidth} $\langle \delta \epsilon \delta \epsilon \rangle$}

\psfrag{XXXXXXXXf0f0lt}[Bl][Br]{\scriptsize\hspace{-0.075\textwidth} $\langle \delta f_0 \delta f_0 \rangle$}
\psfrag{XXXf0f0w}[Bl][Br]{\scriptsize\hspace{-0.075\textwidth} $\langle \delta f_0 \delta f_0 \rangle_H$}
\psfrag{XXXf0f0full}[Bl][Br]{\scriptsize\hspace{-0.075\textwidth} $\langle \delta f_0 \delta f_0 \rangle_f$}

\psfrag{XXXf1f1}[Bl][Br]{\scriptsize\hspace{-0.075\textwidth} $\langle \delta f_1 \delta f_1 \rangle$}
\psfrag{XXXf1f1full}[Bl][Br]{\scriptsize\hspace{-0.075\textwidth} $\langle \delta f_1 \delta f_1 \rangle_f$}
\psfrag{XXXf2f2}[Bl][Br]{\scriptsize\hspace{-0.075\textwidth} $\langle \delta f_2 \delta f_2 \rangle$}
\psfrag{XXXf3f3full}[Bl][Br]{\scriptsize\hspace{-0.075\textwidth} $\langle \delta f_3 \delta f_3 \rangle_f$}
\psfrag{XXXf3f3}[Bl][Br]{\scriptsize\hspace{-0.075\textwidth} $\langle \delta f_3 \delta f_3 \rangle$}
\psfrag{XXXf2f2f4f4full}[Bl][Br]{\scriptsize\hspace{-0.075\textwidth} $\langle \delta f_2 \delta f_2 \rangle_f$}
\psfrag{XXXf4f4}[Bl][Br]{\scriptsize\hspace{-0.075\textwidth} $\langle \delta f_4 \delta f_4 \rangle$}
\psfrag{XXXf5f5}[Bl][Br]{\scriptsize\hspace{-0.075\textwidth} $\langle \delta f_5 \delta f_5 \rangle$}
\psfrag{XXXf5f5f8f8full}[Bl][Br]{\scriptsize\hspace{-0.075\textwidth} $\langle \delta f_5 \delta f_5 \rangle_f$}
\psfrag{XXXf6f6}[Bl][Br]{\scriptsize\hspace{-0.075\textwidth} $\langle \delta f_6 \delta f_6 \rangle$}
\psfrag{XXXf6f6f7f7full}[Bl][Br]{\scriptsize\hspace{-0.075\textwidth} $\langle \delta f_6 \delta f_6 \rangle_f$}
\psfrag{XXXf7f7}[Bl][Br]{\scriptsize\hspace{-0.075\textwidth} $\langle \delta f_7 \delta f_7 \rangle$}
\psfrag{XXXf8f8}[Bl][Br]{\scriptsize\hspace{-0.075\textwidth} $\langle \delta f_8 \delta f_8 \rangle$}

\psfrag{XXXf4}[Bl][Br]{\scriptsize\hspace{-0.05\textwidth} $f_4^0$}
\psfrag{XXXf5}[Bl][Br]{\scriptsize\hspace{-0.05\textwidth} $f_5^0$}

\psfrag{XXXRJX}[Bl][Br]{\scriptsize\hspace{-0.05\textwidth} $\langle \delta \rho \delta j_x \rangle$}
\psfrag{XXXRJY}[Bl][Br]{\scriptsize\hspace{-0.05\textwidth} $\langle \delta \rho \delta j_y \rangle$}
\psfrag{XXXRPXXMPYY}[Bl][Br]{\scriptsize\hspace{-0.05\textwidth} $\langle \delta \rho \delta \Pi_{xx-yy} \rangle$}
\psfrag{XXXRPXY}[Bl][Br]{\scriptsize\hspace{-0.05\textwidth} $\langle \delta \rho \delta \Pi_{xy} \rangle$}
\psfrag{XXXRPXXPPYY}[Bl][Br]{\scriptsize\hspace{-0.05\textwidth} $\langle \delta \rho \delta \Pi_{xx+yy} \rangle$}
\psfrag{XXXRG1}[Bl][Br]{\scriptsize\hspace{-0.05\textwidth} $\langle \delta \rho \delta q_x \rangle$}
\psfrag{XXXRG2}[Bl][Br]{\scriptsize\hspace{-0.05\textwidth} $\langle \delta \rho \delta q_y \rangle$}
\psfrag{XXXRG3}[Bl][Br]{\scriptsize\hspace{-0.05\textwidth} $\langle \delta \rho \delta \epsilon \rangle$}

\psfrag{XXXJXPXXMPYY}[Bl][Br]{\scriptsize\hspace{-0.05\textwidth} $\langle \delta j_x \delta \Pi_{xx-yy} \rangle$}
\psfrag{XXXJXPXXPPYY}[Bl][Br]{\scriptsize\hspace{-0.05\textwidth} $\langle \delta j_x \delta \Pi_{xx+yy} \rangle$}
\psfrag{XXXJYPXY}[Bl][Br]{\scriptsize\hspace{-0.05\textwidth} $\langle \delta j_y \delta \Pi_{xy} \rangle$}
\psfrag{XXXPXXMPYYG1}[Bl][Br]{\scriptsize\hspace{-0.05\textwidth} $\langle  \delta q_x \delta \Pi_{xx+yy} \rangle$}
\psfrag{XXXPXYG2}[Bl][Br]{\scriptsize\hspace{-0.05\textwidth} $\langle \delta \Pi_{xy} \delta q_y \rangle$}
\psfrag{XXXG1G3}[Bl][Br]{\scriptsize\hspace{-0.05\textwidth} $\langle \delta q_x \delta \epsilon \rangle$}

\psfrag{X /   }[Br][Bc]{}
\psfrag{Y /   }[Br][Bc]{}
\psfrag{jx}[Bc][Bc]{\scriptsize$j_x$\normalsize}
\psfrag{qx}[Bc][Bc]{\scriptsize$q_x$\normalsize}
\psfrag{pxy}[Bc][Bc]{\scriptsize$\Pi_{\times}$\normalsize}
\psfrag{rho}[Bc][Bc]{\scriptsize$\rho$\normalsize}
\psfrag{ppl}[Bc][Bc]{\scriptsize$\Pi_{+}$\normalsize}
\psfrag{pmn}[Bc][Bc]{\scriptsize$\Pi_{-}$\normalsize}
\psfrag{eps}[Bc][Bc]{\scriptsize$\epsilon$\normalsize}
\psfrag{jy}[Bc][Bc]{\scriptsize$j_y$\normalsize}
\psfrag{qy}[Bc][Bc]{\scriptsize$q_y$\normalsize}

\begin{document}
\title{A fluctuating lattice-Boltzmann method with improved Galilean invariance}

\author{G.~\surname{Kaehler}}
\email{goetz.kaehler@ndsu.edu}
\author{A.~J.~\surname{Wagner}}
\email{alexander.wagner@ndsu.edu}
\affiliation{Department of Physics, North Dakota State University, Fargo, ND 58108, U.S.A.}

%\emails{{goetz.kaehler@ndsu.edu} (G.~Kaehler)}

\begin{abstract}
In this paper we show that standard implementations of fluctuating Lattice Boltzmann methods do not obey Galilean invariance at a fundamental level. In trying to remedy this we are led to a novel kind of multi-relaxation time lattice Boltzmann methods where the collision matrix depends on the local velocity. This new method is conceptually elegant but numerically inefficient. With a small numerical trick, however, this method recovers nearly the original efficiency and allows the practical implementation of fluctuating lattice Boltzmann methods with significantly improved Galilean invariance. This will be important for applications of fluctuating lattice Boltzmann for non-equilibrium systems involving strong flow fields.
%We investigate the stability and reliability of the now standard lattice-Boltzmann implementations of fluctuating ideal gas \cite{adhikari-2005, duenweg-2007}. In particular we investigate the validity of Galilean invariance. One core assumption of these derivations implies that the transformation matrices involved can be considered velocity independent. It turns out that for large velocities in terms of lattice units this is not generally the case. We introduce a corrections that maintains Galilean invariance to a much better degree.
\end{abstract}
\maketitle

\section{Introduction}
%To do:
%\begin{itemize}
%\item Introduction
%\item Abstract
%\item Lattice Boltzmann Simulation of a fluctuating ideal gas
%\begin{itemize}
%\item Fluctuating LBE $\checkmark$
%\item MRT Introduction $\checkmark$
%\item Adhikari noise introduction $\checkmark$
%\item Hermite Norm $\checkmark$
%\item equilibrium distribution $\checkmark$
%\item $w_i$ approximation $\checkmark$
%\end{itemize}
%\item Galilean Invariance Violations of the hermite norm implementation
%\begin{itemize}
%\item observables (correlators, <dfidfi>, <dmidmi>, 
%\item results for Hermite norm ()
%\end{itemize}
%\item 2nd order orthogonality condition
%\begin{itemize}
%\item New orthogonality condition
%\item locality problem
%\item existance of solutions + corresponding plot
%\item local orthogonalization
%\item computational effort
%\item D2Q9 intro + image
%\item D2Q9 hermite matrices + equilibrium Moments $\checkmark$
%\end{itemize}
%\end{itemize}

Including noise in lattice Boltzmann simulations has been an active field of research in the last few years.
It was pioneered by Ladd\cite{ladd-1993} who suggested to introduce noise on the non-conserved hydrodynamic modes, i.e. the stress modes. This approach works reasonably well in the hydrodynamic limit but for short length scales the fluctuations are underrepresented due to interaction with the non-hydrodynamic degrees of freedom which are typically called the 'ghost'-modes. Adhikari {\em et al.} \cite{adhikari-2005} recognized the necessity to include noise on all non-conserved degrees of freedom, including the non-physical 'ghost'-modes and D\"{u}nweg {\em et al.} \cite{duenweg-2007} reformulated this approach to follow a detailed-balance condition description. All of these publications describe a fluctuating isothermal ideal gas. Just recently there was significant progress in extending this concept to non-ideal equations of state \cite{gross-2010, gross-2011, ollila-2011}.

The Adhikari implementation employs a multi-relaxation time (MRT) method similar to the one originally introduced by d'Humieres \cite{dhumieres-1992} except that the modes are orthogonal with respect to the Hermite norm. This allows for independent relaxation to the physically relevant moments. In particular it simplifies the construction of a noise term that does not violate conservation laws while allowing for non-correlated noise on all other degrees of freedom. The derivation of the fluctuation-dissipation theorem in both, Adhikari's and D\"{u}nweg's approaches requires the MRT transforms to be orthogonal with respect to a certain norm. In the case of a fluctuating ideal gas this norm depends on the equilibrium distribution. However, in all previous publications the equilibrium distribution in this norm is taken only to zeroth order, i.e. only the weight factors in the equilibrium distribution are used. The result is that the MRT orthogonality condition employed is identical to what is typically known as the Hermite norm \cite{benzi-1992}. This approximation, as we first discussed in \cite{kaehler-2011} and show later in this paper, formally introduces non-Galilean invariant terms. We investigate here the effects of using this zeroth order approximation with respect to fluctuations in the context of non-zero flow speeds. The observed Galilean invariance violations suggest that this approximation may be inappropriate in some cases. To avoid this approximation we developed a novel kind of lattice Boltzmann method which includes the full second order expression which we expected to significantly reduce the Galilean invariance violations observed. Such a method necessarily has a local collision matrix that depends on the velocity at the respective lattice site.

% and propose to include the full second order expression to significantly reduce the Galilean invariance violations. 

The paper is structured as follows: In section two we present a more detailed derivation based on Adhikari's noise implementation to show where the non-Galilean invariant terms originate. We elaborate on the source of the orthogonality condition and the consequences of the zeroth order approximation and illustrate the impact on the MRT transforms. In section three we test the current literature standard for the example of a D2Q9 simulation. We measure the validity of two core assumptions of the derivation in the context of large flow speeds and find that Galilean invariance is indeed violated. Section four then discusses approaches to remedy the Galilean invariance violations. In particular we move away from the zeroth order orthogonality condition and attempt to introduce first and second order velocity terms of the equilibrium distribution. As a consequence we derive a lattice Boltzmann method for which the MRT transforms become locally velocity dependent. However, a simplistic implementation of this method is numerically inefficient. This inefficiency can be overcome by introducing look-up tables. The resulting LB scheme's computational cost is only slightly larger than that of the Hermite norm implementation and Galilean invariance violations are significantly reduced.

\section{Lattice Boltzmann simulation of a fluctuating ideal gas}
In order illustrate the origin of Galilean invariance violations in fluctuating lattice Boltzmann implementations we present a short derivation of the fluctuating ideal gas in the Lattice Boltzmann context. The derivation presented is based on Adhikari {\em et al.}'s work \cite{adhikari-2005} who first recognized the necessity to include noise on all non-conserved degrees of freedom. The derivation given in Adhikari {\em et al.}'s original paper is not very detailed and we clarify some of the omitted steps of their derivation in this section. We put emphasis on a clear notation that separates the velocity space distibution functions $f_i$ and the moment space moments we call $M^a$.

The fluctuating lattice-Boltzmann equation is given by
\begin{align}
\label{eqn:LBE1}
&f_i(\vecc{x}+v_i, t+1) = \\\nonumber 
&f_i(\vecc{x}, t) + \sum_j \Lambda_{ij}\left\lbrack f_j(\vecc{x}, t) - f_j^0(\vecc{x},t)\right\rbrack + \xi_i(\vecc{x}, t), 
\end{align}
where the $f_i$ are densities associated with the velocities $v_{i}$. The local equilibrium distribution depends on position and time through the local density $\rho = \sum_i f_i$ and velocity $\mathbf{u} = \sum_i f_i \mathbf{v}_i / \rho$. The structure of the collision matrix $\Lambda_{ij}$ is discussed later in this section. This is the standard BGK lattice-Boltzmann equation with an added noise term $\xi_i(\vecc{x}, t)$. These noise terms must be chosen such that conserved quantities $\rho$, $\vecc{j}$, where $\mathbf{j} = \sum_i f_i \mathbf{v}_i$, are not changed and a proper fluctuation dissipation theorem (FDT) is obeyed. How we obtain the latter while ensuring the former is outlined below.

Throughout this paper we use Qian's second order expansion \cite{qian-1992} of the continuous Maxwell-Boltzmann distribution as expression for the equilibrium distribution
\beq{eqn:f0} 
f_i^0(\rho, \vecc{u}, \theta) = \rho w_i \left\lbrack 1 + \frac{1}{\theta} \vecc{u}.v_i + \frac{1}{2\theta^2}\left(\vecc{u}.v_i\right)^2 - \frac{1}{2\theta}\vecc{u}.\vecc{u}\right\rbrack.
\eeq
This form is typically used for simulations of isothermal hydrodynamics. The extention to thermal hydrodynamics is conceptually straight forward.
All references below to zeroth, first or second order terms in velocity of the equilibrium distribution are to be understood in terms powers of $\vecc{u}$ of this expression.

In order to gain independent access of conserved and non-conserved moments it is useful to shift from Boltzmann type particle distributions $f_i$ to what is called generalized lattice-Boltzmann, moment space representation, or multi relaxation time representation (MRT)\cite{dhumieres-1992, lallemand-2000}. One thus gains access to the hydrodynamically relevant moments directly. For this purpose a set of a forward transform from velocity space and its density functions $f_i$ to moment space and its so-called moments $M^a$
\begin{equation}
\label{eqn:transform1}
M^a(\vecc{x}, t) = \sum_i m_i^a f_i(\vecc{x}, t).
\end{equation}
 and the corresponding back transform
\begin{equation}
\label{eqn:transform2}
f_i(\vecc{x}, t) = \sum_a n_i^a M^a(\vecc{x}, t).
\end{equation}
must be chosen. While the original matrix elements $m_i^a$ and $n_i^a$ in \cite{dhumieres-1992} were identical this is not necessary. But they need to follow the orthogonality conditions
\begin{equation}
\label{eqn:symmetry}
\sum_i m_i^a n_i^b = \delta^{ab} \text{ and } \sum_a m_i^a n_j^a = \delta_{ij}.
\end{equation}
The particular choice of these transforms aims to generate a simple form for the fluctuation dissipation theorem and is of key importance to the validity of the noise derivation and Galilean invariance or lack thereof. As such they differ from those in the publications introducing the MRT formalism \cite{dhumieres-1992, lallemand-2000}. At least in the case of the ideal gas implementation it is convenient to choose the moments $M^a$ such that the representation of the collision matrix $\Lambda$ in moment space is diagonal $\Lambda^{ab} = \frac{t}{\tau^a} \delta^{ab}$.
For practical purposes it is then useful to perform the collision in moment space. The fluctuating LBE \eref{eqn:LBE1} is then written as
%\bal{eqn:LBEMRT}
\begin{align}
\label{eqn:LBEMRT}
&f_i(\vecc{x} + v_i, t + 1) - f_i(\vecc{x}, t) = \\
\nonumber 
&\sum_a n_i^a \lbc\sum_b \Lambda^{ab} \lbk M^b(\vecc{x}, t) - M^{b,0}(\vecc{x},t)\rbk + \xi^a N \rbc
\end{align}
%\eal
where $\xi^a$ is the noise amplitude associated with moment $M^a$ and $N$ is a random number chosen from a Gaussian distribution with a variance of one. The primary advantage here is that we gain independent access to the hydrodynamically relevant physical moments and we can choose the noise amplitudes $\xi^a$ such that conservation laws are not violated, i.e. $\xi^{a, conserved} = 0$.

Now we separate the $f_i$ in \eref{eqn:LBE1} into their global mean values and a local fluctuating term 
\beq{eqn:dfdef}
f_i = \langle f_i \rangle + \delta f_i
\eeq
 and we obtain
%\beq{eqn:LBE2}
\begin{align}
\label{eqn:LBE2}
&\langle f_i \rangle + \delta f_i(\vecc{x}+v_i, t+1) = \langle f_i \rangle + \delta f_i(\vecc{x}, t) \\ \nonumber
&+ \sum_j \Lambda_{ij}\left\lbrack \langle f_j \rangle + \delta f_j(\vecc{x}, t) - \langle f_j^0 \rangle - \delta f_j^0(\vecc{x},t)\right\rbrack \\ \nonumber
&+ \xi_i(\vecc{x}, t).
\end{align}
%\eeq
Subtracting the $\langle f_i \rangle$ and assuming 
\beq{eqn:efif0} 
\langle f_i \rangle = f_i^0(\rho_0, \mathbf{u}_0),
\eeq  
where $\rho_0$ and $\mathbf{u}_0$ are the equilibrium values of the density and the velocity, yields a LBE for the fluctuation part of the distribution
\begin{align}
\label{eqn:LBE3}
&\delta f_i(\vecc{x}+v_i, t+1) = \\ \nonumber
&\delta f_i(\vecc{x}, t) + \sum_j \Lambda_{ij}\left\lbrack \delta f_j(\vecc{x}, t) - \delta f_j^0(\vecc{x},t)\right\rbrack + \xi_i(\vecc{x}, t).
\end{align}
We can now Fourier transform in space and apply the moment space transform $\sum_i m_i^a$ to obtain the moment space evolution equation in $k$-space
\begin{align}
\label{eqn:LBEft5}
&\delta M^a(k, t+1) = \sum_i \sum_b m_i^a e^{-ikv_i} n_i^b \Big\lbrace \delta M^b(k, t) + \\\nonumber
&\sum_j \sum_c \sum_d \Lambda^{bc} m_j^c n_j^d \left\lbrack \delta M^d(k,t) - \delta M^{0,d}(k,t)\right\rbrack +\\\nonumber
&\xi^b(k,t)\Big\rbrace,
\end{align}
where we also used $\Lambda_{ij} = \sum_a\sum_b n_i^a \Lambda^{ab} m_j^b$.
We now assume that we can choose the moments such that the multi relaxation time collision operator is diagonal in moment space, i.e. $\Lambda^{ab} = -\delta^{ab} \frac{1}{\tau^a}$. Using $\Gamma^{ab}(k) = \sum_i m_i^a n_i^b e^{-ikv_i}$ and $\delta M^0 = 0$ we thus get the evolution equation of the fluctuations in spatial Fourier representation of moment space
\begin{align}
\label{eqn:LBEft8}
&\delta M^a(k, t+1) = \\\nonumber
&\sum_b \Gamma^{ab}(k) \left\lbrace \left(1 - \frac{1}{\tau^b} \right) \delta M^b(k,t) +\xi^b(k,t)\right\rbrace.
\end{align}

Taking the outer product of $\delta M^a$ with itself, performing an ensemble average and substituting $r^a = 1-1/\tau^a$ we obtain
\begin{align}
\label{eqn:outerproduct}
&\left\langle \delta M^a(k, t+1) \delta M^c(k, t+1) \right\rangle = \\\nonumber 
&\Big\langle \sum_b \sum_d \Gamma^{ab} \big\lbrack r^b \delta M^b(k, t) + \xi^b \big\rbrack \Gamma^{cd}\\\nonumber 
&\left\lbrack r^d \delta M^d(k, t) + \xi^d \right\rbrack \Big\rangle.
\end{align}
For an ideal gas we know the results to be $\vecc{k}$-independent. Henceforth Adhikari {\em et al.} only consider the case $k = 0$ at which $\Gamma^{ab} = \delta^{ab}$. They also invoke stationarity of equal time correlators $\langle \delta M^a (t+1) \delta M^b (t+1) \rangle  = \langle \delta M^a(t) \delta M^b(t)\rangle$ and get
\begin{align}
\label{eqn:outerproduct2}
&\left\langle \delta M^a(t+1) \delta M^c(t+1) \right\rangle = r^c r^a \left\langle \delta M^a(t) \delta M^c(t) \right\rangle + \\ \nonumber 
&r^c \left\langle \delta M^c(t) \xi^a(t) \right\rangle + r^a \left\langle \delta M^a(t) \xi^c(t) \right\rangle + \left\langle \xi^a \xi^c \right\rangle.
\end{align}
Now, using the fact that the current system state is independent of the noise contribution, i.e. $\langle \delta M^a \xi^a \rangle = 0$, they obtain
\begin{eqnarray}
\label{eqn:FDT}
\left\langle \xi^a \xi^c \right\rangle & = & (1 - r^a r^c) \left\langle \delta M^a \delta M^c \right\rangle \nonumber \\
& = & \frac{\tau^a + \tau^c - 1}{\tau^a \tau^c} \left\langle \delta M^a \delta M^c \right\rangle ,
\end{eqnarray}
which acts as the fluctuation dissipation theorem (FDT). It relates the noise to the moment fluctuations. What is left is finding a prediction for $\langle \delta M^a \delta M^b \rangle$.
Assuming the case of the ideal gas ~\cite{lifshitz-1981} they use the fact that the distribution functions $f_i$ follow Poisson statistics with a mean value and variance of $\langle f_i\rangle$. Thus with \eref{eqn:efif0} they get 
\beq{eqn:dfdf}\langle \delta f_i \delta f_j \rangle = f_i^0 \delta_{ij}.
\eeq
The back transform to velocity space can now be applied to the moment space correlator to obtain
\begin{align}
\label{eqn:fluctf2}
\langle \delta M^a \delta M^b \rangle = \sum_i \sum_j m_i^a m_j^b \langle \delta f_i \delta f_j\rangle = \\ \nonumber \sum_i \sum_j m_i^a m_j^b f_i^0 \delta_{ij}.
\end{align}
This implies that the moment fluctuations and by \eref{eqn:FDT} the noise terms are generally correlated. However, we can decouple these terms by choosing $n_i^a = m_i^a f_i^0/\rho$ because then according to \eref{eqn:symmetry} 
\beq{eqn:fluctf3}
\sum_i m_i^a m_i^b f_i^0/\rho = \delta^{ab}
\eeq
and thus
\beq{eqn:fluctf4}
\langle \delta M^a \delta M^b \rangle = \rho \delta^{ab}. 
\eeq
Of course one has also to show that this is also consistent with identifying the $M^a$ with the hydrodynamic moments. For a discussion of this see \cite{kaehler-2011}.

Now that it has been established that the moment fluctuations can be decoupled according to \eref{eqn:fluctf4} we can solve \eref{eqn:FDT} for the noise amplitude
\begin{equation}
\label{eqn:noiseamp}
\xi^a = \frac{1}{\tau^a} \sqrt{\rho \left(2 \tau^a - 1\right)} .
\end{equation}

The actual implementation performes the collision in moment space according to \eref{eqn:LBEMRT} where the moments $M^b$ are constructed at each time step by the standard forward transform. The streaming, however, still has to happen in velocity space and consequently each update involves two matrix transforms. 

Of course, the problem here is that such an orthogonality condition \eref{eqn:fluctf3} is difficult to fulfill at all times and it is not entirely clear which values for $\rho$ and $\vecc{u}$ we have to choose for use in the equilibrium distribution. Both Adhikari\cite{adhikari-2005} and D\"unweg\cite{duenweg-2007} implicitly assume very low flow speeds or the zeroth order expression
\beq{eqn:u=0}
\lim_{\vecc{u} \to 0} f_i^0\lpar\rho, \vecc{u}\rpar = \rho w_i ,
\eeq
thereby avoiding aforementioned problem and simplifying the orthogonality condition to
\beq{eqn:fluctf5}
\sum_i m_i^a m_i^b w_i = \delta^{ab}.
\eeq
This implies $n_i^a = m_i^a w_i$ and is identical to what is frequently called the Hermite norm and was originally introduced by Benzi \cite{benzi-1992}. The orthogonality condition \eref{eqn:fluctf5} therefore qualifies the requirements on the transforms in addition to the necessity that they preserve hydrodynamics. An extensive study on the second condition has been published in \cite{kaehler-2011}. There we found that the Hermite norm of \eref{eqn:fluctf5}, does indeed also preserve hydrodynamics and that, in fact, we are free to add any conserved quantity moments to hydrodynamic modes without impacting the validity of the hydrodynamic equations.
The choice of the zeroth order approximation in \eref{eqn:fluctf5} is, however, not well documented or motivated in the original literature and gives rise to the question whether Galilean invariance violations of the fluctuations result as a consequence.

\section{Galilean Invariance Violations in the Hermite Norm Implementation}

\begin{figure}
\subfigure{
\begin{psfrags} % Optional
\includegraphics[width=0.2\textwidth, angle = 0]{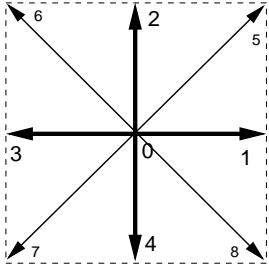}
\end{psfrags}
}
% Optional
\caption{Basis vectors $v_i$ of the D2Q9 scheme used in all simulations in this manuscript.}
\label{fig:d2q9}
\end{figure}

First we want to evaluate what effect choosing the simplified norm of \eref{eqn:fluctf5} has on the Galilean invariance of a fluctuating lattice Boltzmann implementation. Here we show the numerical results for an isothermal D2Q9 fluctuating lattice Boltzmann method with periodic boundary conditions. Moment space transforms are generated with respect to the Hermite norm of \eref{eqn:fluctf5}. The basis vectors $v_i$ are shown in Fig.~\figref{fig:d2q9}. All $i$ indices in the following correspond to these basis vectors. The details of the D2Q9 Hermite norm transforms and the equilibrium moments are documented in appendix \sref{sec:apphermite}.

The results in the following were all obtained in a 2D lattice Boltzmann simulation of size $21\times21$. The odd side lengths are chosen to avoid the independent conservation of momentum components in odd and even lattice sites in either dimension. They occur for even side lengths because collisions conserve momentum and streaming of the densities that constitute momentum and could interact always moves two lattice sites at once. Consequently momenta in odd and even numbered lattice sites would never interact. We use a large average density of $\rho_0 = 10^6$ to avoid stability issues due to local negative density events. These can occur when the noise $\xi_i$ on the distribution functions $f_i$ exceeds the value of these distribution functions. This is more likely for small $\rho$ as the noise amplitude in moment space \eref{eqn:noiseamp} is proportional to $\sqrt{\rho}$. 
All averages were taken over a simulation time of $10^6$ iterations after a thermalization phase of $10^5$ iterations to equilibrate the system.

%because we use a Gaussian distribution for the random noise $\xi^a N$ in \eref{eqn:LBEMRT} and negative moment can happen if the amplitude \eref{eqn:noiseamp} is larger than the local density $\rho(x, y)$. 

The fundamental identity that allows us to decouple the moment fluctuations is given by \eref{eqn:dfdf}. We can verify its validity in the simulation directly by measuring $\langle \delta f_i \delta f_j   \rangle$ as a function of $u_{x,0}$ and comparing it to $f_i^0$ and $w_i$ of \eref{eqn:dfdf} and \eref{eqn:dfidfj2dmadmb}. If the ideal gas hypothesis were to hold we would expect \eref{eqn:dfdf} to be fulfilled independently of $\vecc{u}$. However, using only the Hermite norm \eref{eqn:fluctf5} suggests that we might only find \eref{eqn:dfdf} fulfilled to zeroth order, i.e. to the weight factors $w_i$.

\begin{figure}
\begin{psfrags} % Optional\psfrag{XXXf1f1}[Bl][Br]{\scriptsize\hspace{-0.075\textwidth} $\langle \delta f_1 \delta f_1 \rangle$}
\includegraphics[width=0.3\textwidth, angle = 270]{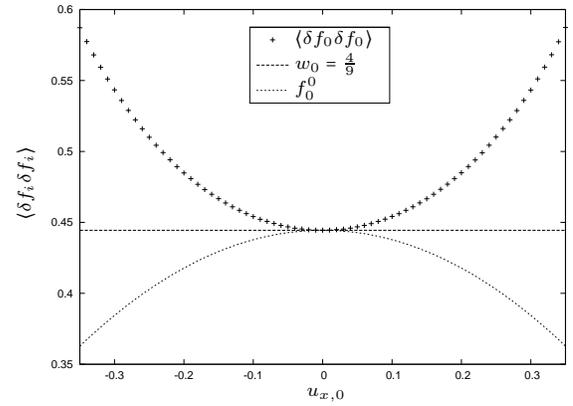}
\end{psfrags}
% Optional
\caption{$\langle \lpar \delta f_0 \rpar^2 \rangle$ in a $21\times21$ D2Q9 fluctuating LB simulation employing the Hermite norm. We plot $w_i$ and $f_i^0$for comparison.}
\label{fig:wdf0df0}
\end{figure}

\begin{figure}
\begin{psfrags} % Optional
\includegraphics[width=0.3\textwidth, angle = 270]{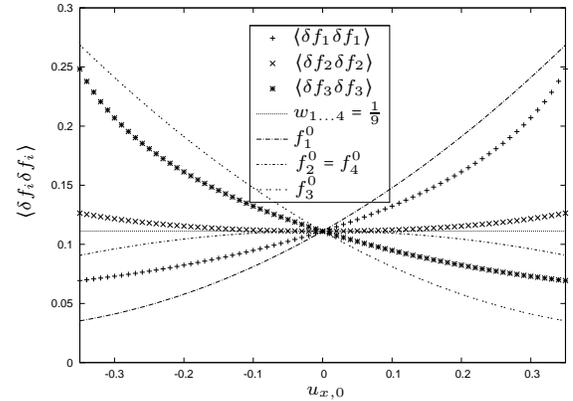}
\end{psfrags}
% Optional
\caption{$\langle  \lpar \delta f_i \rpar^2 \rangle$ for $i=1...3$ in a $21\times21$ D2Q9 fluctuating LB simulation employing the Hermite norm. We plot $w_i$ and $f_i^0$ for comparison. $\langle  \lpar \delta f_4 \rpar^2 \rangle$ is not shown as it is identical to $\langle  \lpar \delta f_2 \rpar^2 \rangle$ for symmetry reasons.}
\label{fig:wdf14df14}
\end{figure}

\begin{figure}
\begin{psfrags} % Optional
\includegraphics[width=0.3\textwidth, angle = 270]{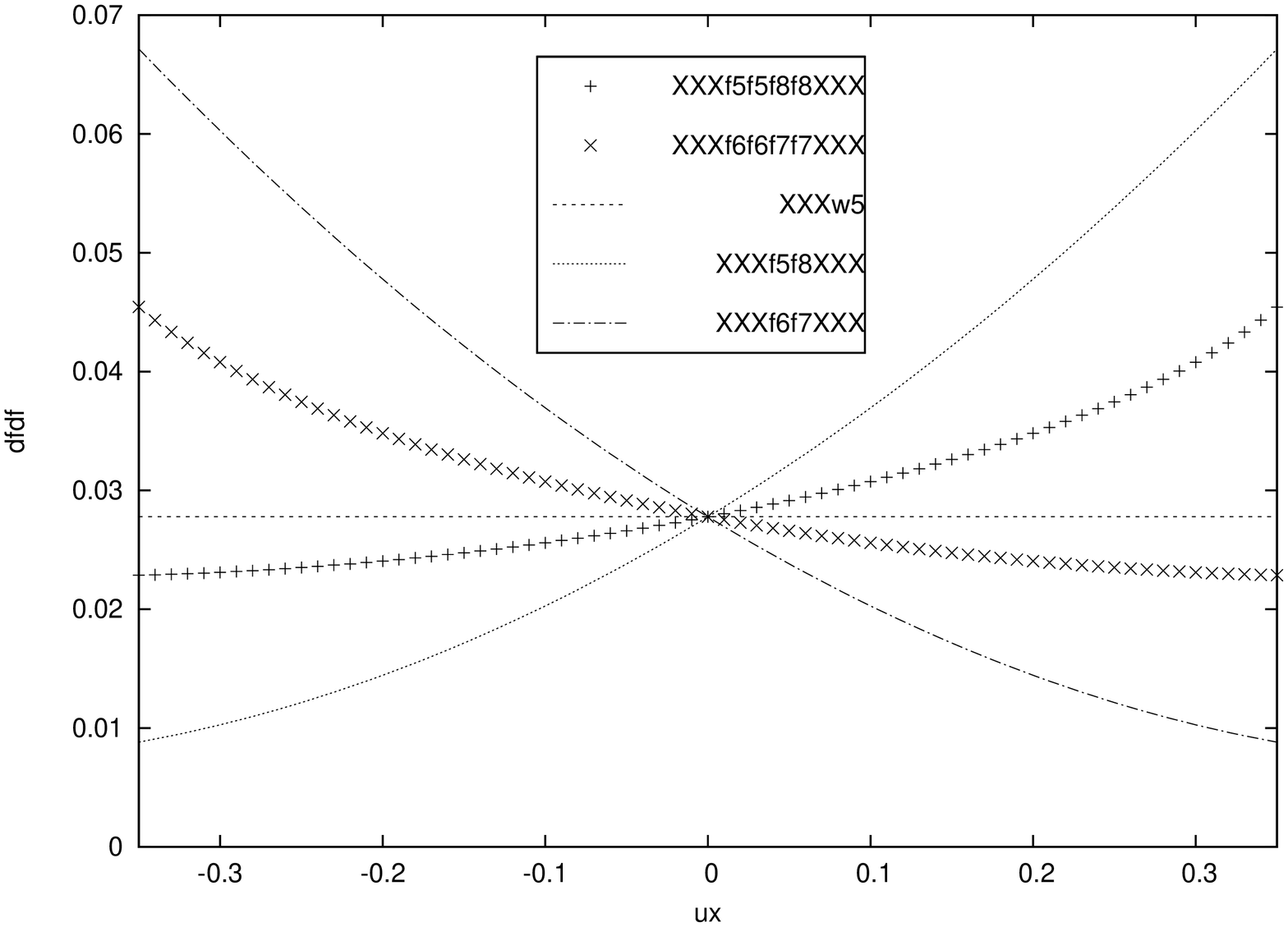}
\end{psfrags}
% Optional
\caption{$\langle \lpar \delta f_i \rpar^2 \rangle$ for $i=5...8$ in a $21\times21$ D2Q9 fluctuating LB simulation employing the Hermite norm. We plot $w_i$ and $f_i^0$ for comparison. $\langle \lpar \delta f_8 \rpar^2 \rangle$ and $\langle \lpar \delta f_7 \rpar^2 \rangle$ are not shown as they appears identical to $\langle \lpar \delta f_5 \rpar^2 \rangle$ and $\langle \lpar \delta f_6 \rpar^2 \rangle$ respectively in the scale of this plot.}
\label{fig:wdf58df58}
\end{figure}

\begin{figure}
\begin{psfrags} % Optional
\includegraphics[width=0.3\textwidth, angle = 270]{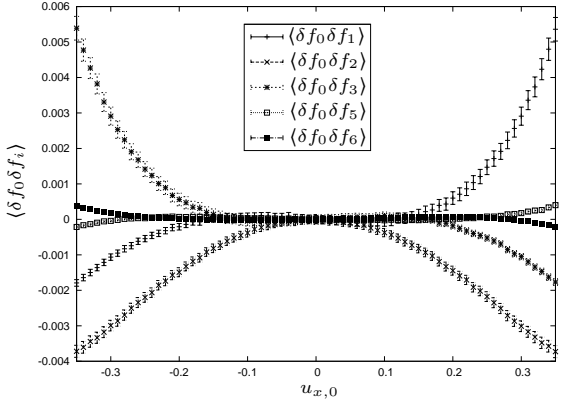}
\end{psfrags}
% Optional
\caption{Off-diagonal correlators $\langle \delta f_0 \delta f_i  \rangle$ for $i=1...8$ in a $21\times21$ D2Q9 fluctuating LB simulation employing the Hermite norm. $\langle \delta f_0 \delta f_4 \rangle$, $\langle \delta f_0 \delta f_7 \rangle$, and $\langle \delta f_0 \delta f_8 \rangle$ are omitted as they behave identical to $\langle \delta f_0 \delta f_2 \rangle$, $\langle \delta f_0 \delta f_6 \rangle$, and $\langle \delta f_0 \delta f_5 \rangle$ respectively.}
\label{fig:wdf0dfi}
\end{figure}

In Figs.~\ref{fig:wdf0df0},~\ref{fig:wdf14df14},~\ref{fig:wdf58df58} we show the simulation results of all unique $\langle \delta f_i \delta f_i \rangle$ correlators as functions of $u_{x,0}$. We find that with increasing velocity $u_{x,0}$ we do indeed deviate strongly from both, the weights $w_i$, and the equilibrium distributions $f_i^0$. In this implementation the correlators approach neither the $w_i$ nor the $f_i^0$ and in some cases not even an intermediate value. For correlators corresponding to base velocities without an $x$-component ($\langle \delta f_0^2\rangle$, $\langle \delta f_2^2\rangle$, $\langle \delta f_4^2\rangle$) the trend opposes that of the  $f_i^0$. In these plots and all similar figures in this paper the statistical error bars are omitted in the graphs when they are smaller than the symbol size.

\begin{figure}
\begin{psfrags} % Optional

\psfrag{XXXXXXXXXXXXXXXXXRR}[Bl][Br]{\scriptsize\hspace{-0.1\textwidth} $\langle \delta \rho \delta\rho \rangle$}
\psfrag{XXXJXJX}[Bl][Br]{\scriptsize\hspace{-0.10\textwidth} $\langle \delta j_x \delta j_x \rangle$}
\psfrag{XXXJYJY}[Bl][Br]{\scriptsize\hspace{-0.10\textwidth} $\langle \delta j_y \delta j_y \rangle$}
\psfrag{XXXPXXMPYY}[Bl][Br]{\scriptsize\hspace{-0.1\textwidth} $\langle \lpar \delta \Pi_{xx-yy} \rpar ^2 \rangle$}
\psfrag{XXXPXY}[Bl][Br]{\scriptsize\hspace{-0.10\textwidth} $\langle \lpar \delta \Pi_{xy} \rpar^2 \rangle$}
\psfrag{XXXPXXPPYY}[Bl][Br]{\scriptsize\hspace{-0.10\textwidth} $\langle \lpar \delta \Pi_{xx+yy} \rpar^2 \rangle$}
\psfrag{XXXG1G1}[Bl][Br]{\scriptsize\hspace{-0.10\textwidth} $\langle \delta q_x \delta q_x \rangle$}
\psfrag{XXXG2G2}[Bl][Br]{\scriptsize\hspace{-0.10\textwidth} $\langle \delta q_y \delta q_y \rangle$}
\psfrag{XXXG3G3}[Bl][Br]{\scriptsize\hspace{-0.10\textwidth} $\langle \delta \epsilon \delta \epsilon \rangle$}

\includegraphics[width=0.3\textwidth, angle = 270]{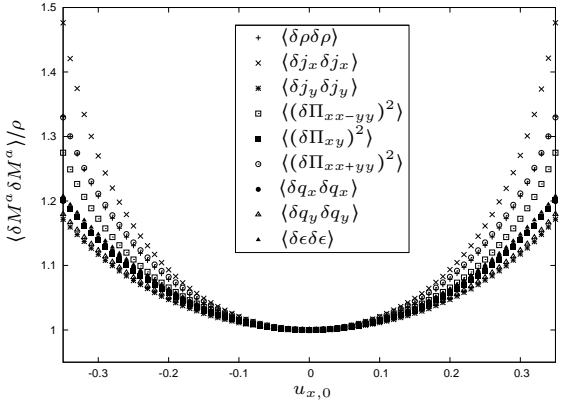}
\end{psfrags}
% Optional
\caption{Correlators calculated in the Hermite norm $\langle \delta M^a \delta M^a\rangle$ normalized to $\rho$ according to \eref{eqn:fluctf4} in a $21\times21$ D2Q9 fluctuating LB simulation employing the Hermite norm.}
\label{fig:wdmadma}
\end{figure}

%\begin{figure}
%\label{fig:wdmadm0}
%\begin{psfrags} % Optional
%\includegraphics[width=0.3\textwidth, angle = 270]{wi_m0ma.eps}
%\end{psfrags}
% Optional
%\caption{Correlators $\langle \delta M^0 \delta M^b\rangle$ for $a=1...8$ normalized to $\rho$ according to% \eref{eqn:fluctf4} in a $21\times21$ D2Q9 fluctuating LB simulation employing the Hermite norm.}
%\end{figure}

%If we had $\langle \delta f_i \delta f_j \rangle = f_i^0 \delta_{ij}$ fulfilled, 
In previous publications \cite{adhikari-2005, gross-2010} the fluctuations were characterized by the fluctuations of the hydrodynamics and ghost moments. The corresponding moment correlators follow directly from the distribution function deviations according to 
\beq{eqn:dfidfj2dmadmb}
\langle \delta M^a \delta M^b \rangle = \sum_{ij} m_i^a m_j^b \langle \delta f_i \delta f_j \rangle.
\eeq
and are arguably of more practical importance since they represent the fluctuations of the hydrodynamic fields.

These correlators were expected, in the theory of \cite{adhikari-2005, duenweg-2007, gross-2010, gross-2011, ollila-2011} to obey $\langle \delta M^a \delta M^b \rangle = \rho \delta_{ab}$. However, for this to work we would need $\langle \delta f_i \delta f_j \rangle = w_i$ in \eref{eqn:dfidfj2dmadmb}, which is not the case for non-zero velocities, as we have shown above. We show the observed deviations for the diagonal correlators in Fig.~\figref{fig:wdmadma}. Here the correlator of the current in $x$-direction, $\langle \delta j_x \delta j_x \rangle$, exhibits the largest deviations.

%This now allows us to write down what $\langle \delta M^a \delta M^b \rangle$ looked like if $\langle \delta f_i \delta f_j \rangle = w_i \delta_{ij}$. With \eref{eqn:u=0} we then find $\langle \delta M^a \delta M^b \rangle = \rho \delta^{ab}$. If we had instead \eref{eqn:fluctf2} we would get the distribution functions shown in Figs.~\ref{fig:wdf0df0},~\ref{fig:wdf14df14},~\ref{fig:wdf58df58}.
%However, as they do not and because we need to relate the LB method to hydrodynamics, we focus on the correlators of the physical moments $M^a$ in addition to those of the $f_i$. These can be measured \eref{eqn:fluctf4} directly. They, again, should hold independently of the fluid velocity $\vecc{u}$. We find that with increasing velocity $\vecc{u}$ we find significant deviations from \eref{eqn:fluctf4}. All of the square correlators deviate as seen in Fig.~\figref{fig:wdmadma}, the most pronounced being $\langle j_x j_x \rangle$. 

Note that, while most $f_i$ are not symmetric with regard to the $u_{x,0} \rightarrow -u_{x,0}$ inversion, all the moments are constructed to be either symmetric or antisymmetric under $u_{x,0} \rightarrow -u_{x,0}$.
%The off diagonal correlators $\rangle \delta \rho \delta M^b \langle$ in Fig.~\figref{fig:wdm0ma} again show deviations with increasing $\vecc{u}$. 

\begin{figure}
\subfigure[Linear coefficient $l$, Hermite norm]{
\begin{psfrags} % Optional
\includegraphics[width=0.3\textwidth, angle = 0]{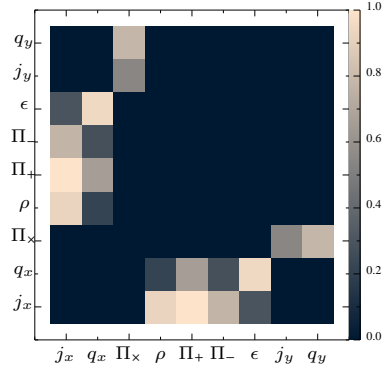}
\end{psfrags}
}
%\subfigure[Linear coefficient $l$, $f$-norm]{
%\begin{psfrags} % Optional
%\includegraphics[width=0.3\textwidth, angle = 0]{fi_m_matrix1.eps}
%\end{psfrags}
%}
\subfigure[Quadratic coefficient $q$, Hermite norm]{
\begin{psfrags} % Optional
\includegraphics[width=0.3\textwidth, angle = 0]{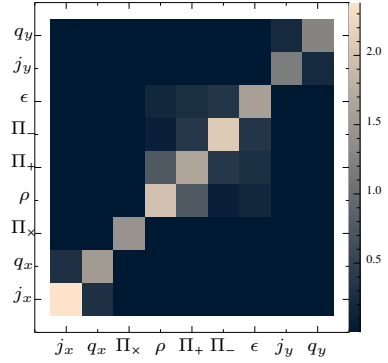}
\end{psfrags}
}
%\subfigure[Quadratic coefficient $q$, $f$-norm]{
%\begin{psfrags} % Optional
%\includegraphics[width=0.3\textwidth, angle = 0]{fi_m_matrix2.eps}
%\end{psfrags}
%}
% Optional
\caption{Linear and quadratic coefficient $l$ and $q$ of all $81$ ($45$ unique) correlators as a result of fitting $\langle \delta M^a \delta M^b\rangle(u_{x,0})-\delta^{ab}$ to $l u_{x,0} + q u_{x,0}^2$. Brighter color indicates larger coefficients. Moments were reordered to visually identify correlations better. To accommodate for symbol size the stress moments were simplified: $\Pi_\times = \Pi_{xy}, \Pi_- = \Pi_{xx-yy}, \Pi_+ = \Pi_{xx+yy}$). The coefficient at position (0, 1) in image (a) would correspond to linear portion of the $\langle \delta j_x \delta q_x \rangle$ correlator. 
Coefficients were measured on a $21 \times 21$ D2Q9 simulation employing the Hermite norm. Fit range used was $-0.25 <= u_x <= 0.25 $.
%for (a) and (c) and the $f$-norm with look up tables,  $u_g = 0.02$, for (b) and (d). Fit range used was $-0.25 <= u_x <= 0.25 $. We observe a significant decrease in both the linear and quadratic coefficients when employing the $f$-norm.
}
\label{fig:wdmadmbfit1}
\end{figure}

%
%\begin{figure}
%\label{fig:wdmadmb}
%\begin{psfrags} % Optional
%\includegraphics[width=0.3\textwidth, angle = 270]{wi_mamb.eps}
%\end{psfrags}
% Optional
%\caption{$\langle \delta M^a \delta M^b\rangle(u_x)$ with the strongest deviations.In particular we notice that $\langle \Pi_{xx-yy} q_x \rangle$ anticorrelate.}
%\end{figure}
%

To obtain some quantitative measure of the dependency of all 81 (45 unique) correlators in 
\eref{eqn:dfidfj2dmadmb} we fit a second order polynomial $l u_{x,0} + q u_{x,0}^2$ to $\langle \delta M^a \delta M^b \rangle / \rho_0 - \delta^{ab}$. The resulting coefficients $l$ for odd combinations and $q$ for even combinations give a rough estimate of the deviation of the particular moment correlators and are depicted in Fig.~\figref{fig:wdmadmbfit1}. We notice in Fig.~\figref{fig:wdmadmbfit1}(b) that while the quadratic dependency of the correlations on the velocity is present in several correlators, it is particularly apparent on the square correlators. The linear dependency only appears in cross-correlators which are anti-symmetric under $u_{x,0} \rightarrow -u_{x,0}$ as seen in Fig.~\figref{fig:wdmadmbfit1}(a).

%negative sign correlators: 
%linear: 14 (rho q_x), 36(jx epsilon), 68(Pxxmyy j_x), 110, 116, 148
%quadratic: 12 (rho Pxxpyy), 32(j_x q_x), 72(Pxxmyy epsilon), 92(rho Pxxpyy), 108(Pxxpyy epsilon), 112, 152, 156

The ensemble averages of the correlation functions shown so far do not resolve the length scale dependency of the deviations we observed. To gain some understanding here we measure the static structure factor 
\begin{equation}
\label{eqn:sfactorr}
S_{\vecc{k}}(\rho) = \frac{1}{\rho_0} \left\langle \delta \rho(\vecc{k}) \delta \rho(\vecc{-k}) \right\rangle, 
\end{equation}
the $j_x$ momentum correlator
\begin{equation}
\label{eqn:sfactoru}
S_{\vecc{k}}(j_x) = \frac{1}{\rho_0} \left\langle \delta j_x(\vecc{k}) \delta j_x(\vecc{-k}) \right\rangle, 
\end{equation}
at chosen velocities and the momentum cross correlator
\begin{equation}
\label{eqn:rcorrelator}
R_{\vecc{k}}(j_x, j_y) = \frac{1}{\rho_0} \left\langle \delta j_x(\vecc{k}) \delta j_y(\vecc{-k}) \right\rangle
\end{equation}
at imposed average system velocities $u_{x,0} = 0.0$, $u_{x,0} = 0.1$, and $u_{x,0} = 0.2$. We chose $R_{\vecc{k}}(j_x, j_y)$ in reference to Donev {\it et al.}'s investigation of the accuracy of finite volume schemes \cite{donev-2009}.

Here $\delta \rho(\vecc{k}) = \sum_{\vecc{x}} \lbrack\rho(\vecc{x})-\rho_0\rbrack e^{-i \vecc{k} \cdot \vecc{x}}$ and $\delta j_x(\vecc{k}) = \sum_{\vecc{x}} \lbrack j_x(\vecc{x})-j_{x,0} \rbrack e^{-i \vecc{k} \cdot \vecc{x}}$ are the discrete spatial Fourier transforms and $\sum_{\vecc{x}}$ is understood to be the summation over all discrete lattice sites. 

\begin{figure}
\begin{psfrags}
\psfrag{X /   }[Bl][Bc]{$k_x$}
\psfrag{Y /   }[Bl][Bc]{$k_y$}
\psfrag{ 10^-3}[Bl][Bc]{$10^{-3}$}
\subfigure[$u_{x,0}=0.0$]{\includegraphics[width=.3\textwidth]{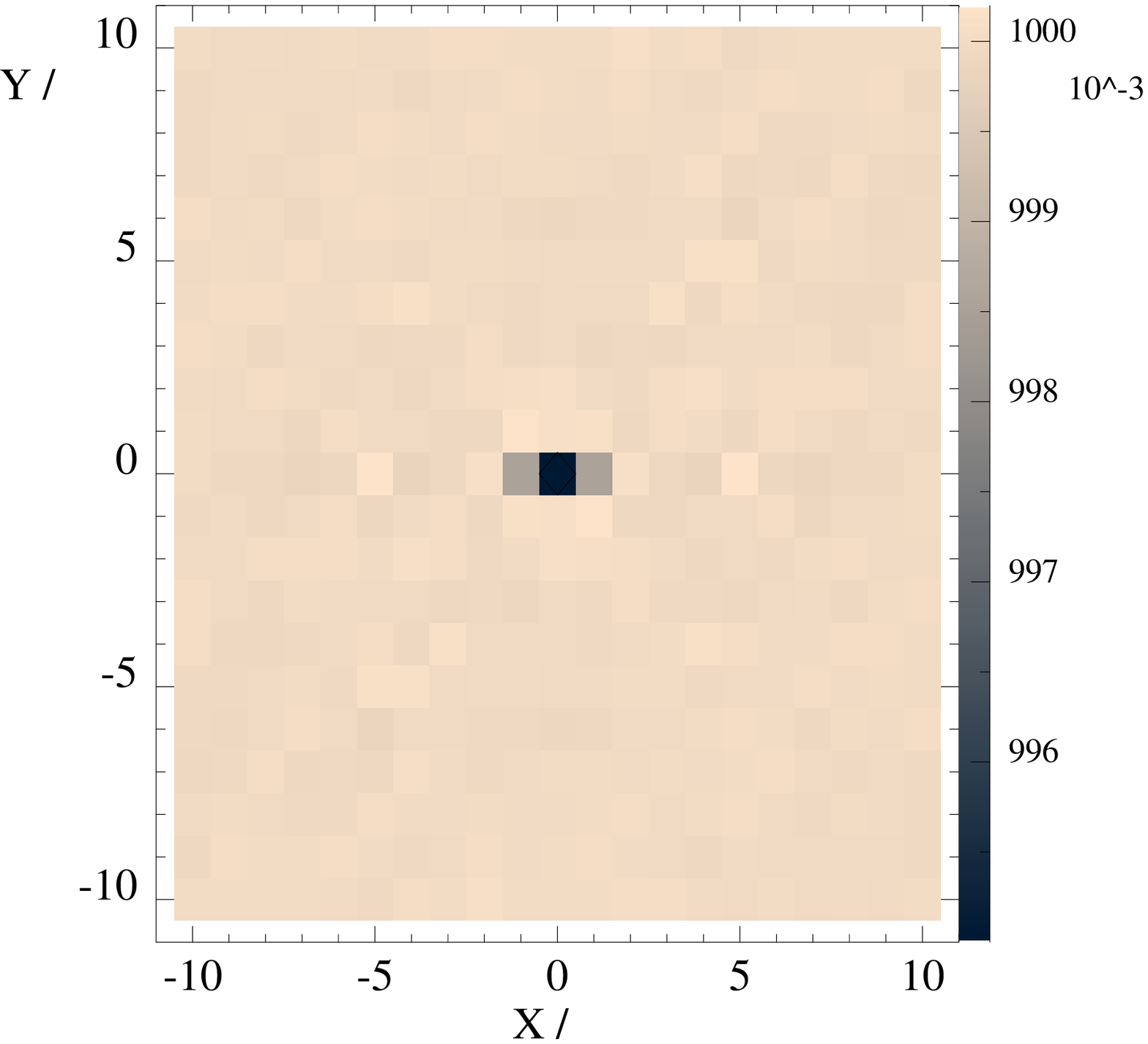}}
%\subfigure[$u_{x,0}=0.1$]{\includegraphics[width=.3\textwidth]{wi_d2q9gurr.eps}}
\subfigure[$u_{x,0}=0.1$]{\includegraphics[width=.3\textwidth]{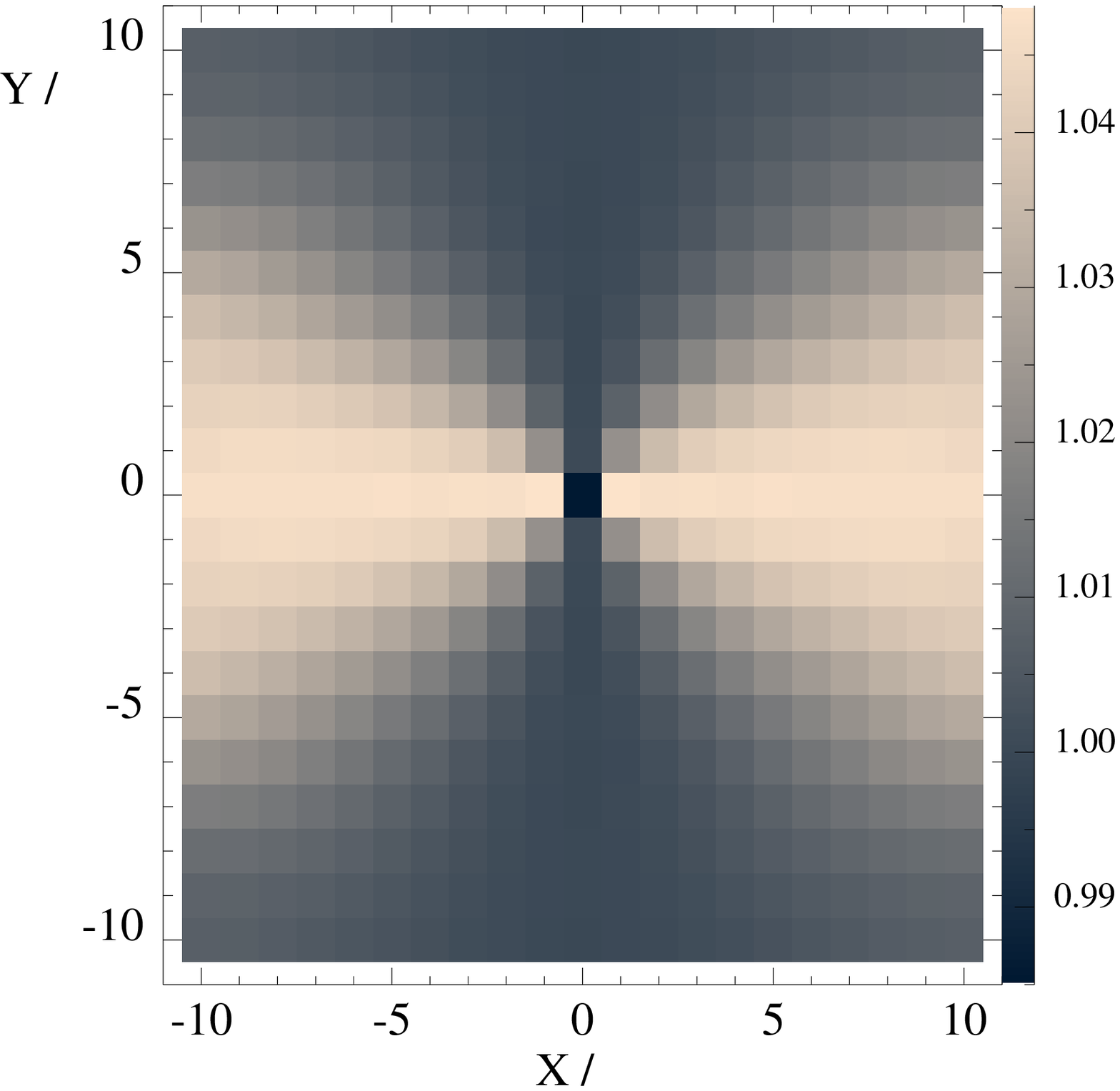}}
\subfigure[$u_{x,0}=0.2$]{\includegraphics[width=.3\textwidth]{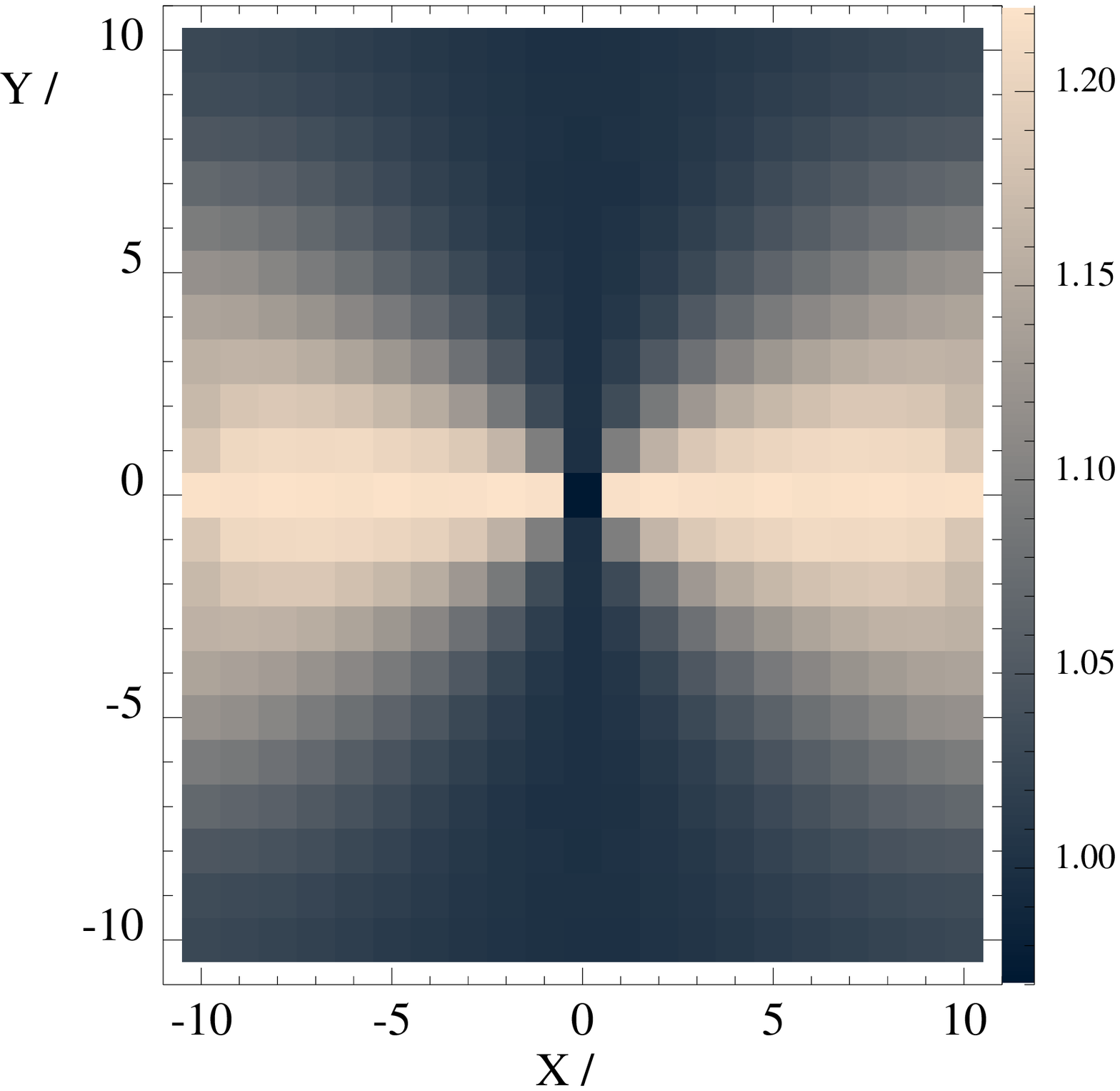}}
\end{psfrags}
\caption{Static structure factor $S_{\vecc{k}}(\rho)$ at different velocities measured for the Hermite norm.}
\label{fig:wsrr}
\end{figure}

\begin{figure}
\begin{psfrags}
\psfrag{X /   }[Bl][Bc]{$k_x$}
\psfrag{Y /   }[Bl][Bc]{$k_y$}
\psfrag{ 10^-3}[Bl][Bc]{$10^{-3}$}
\subfigure[$u_{x,0}=0.0$]{\includegraphics[width=.3\textwidth]{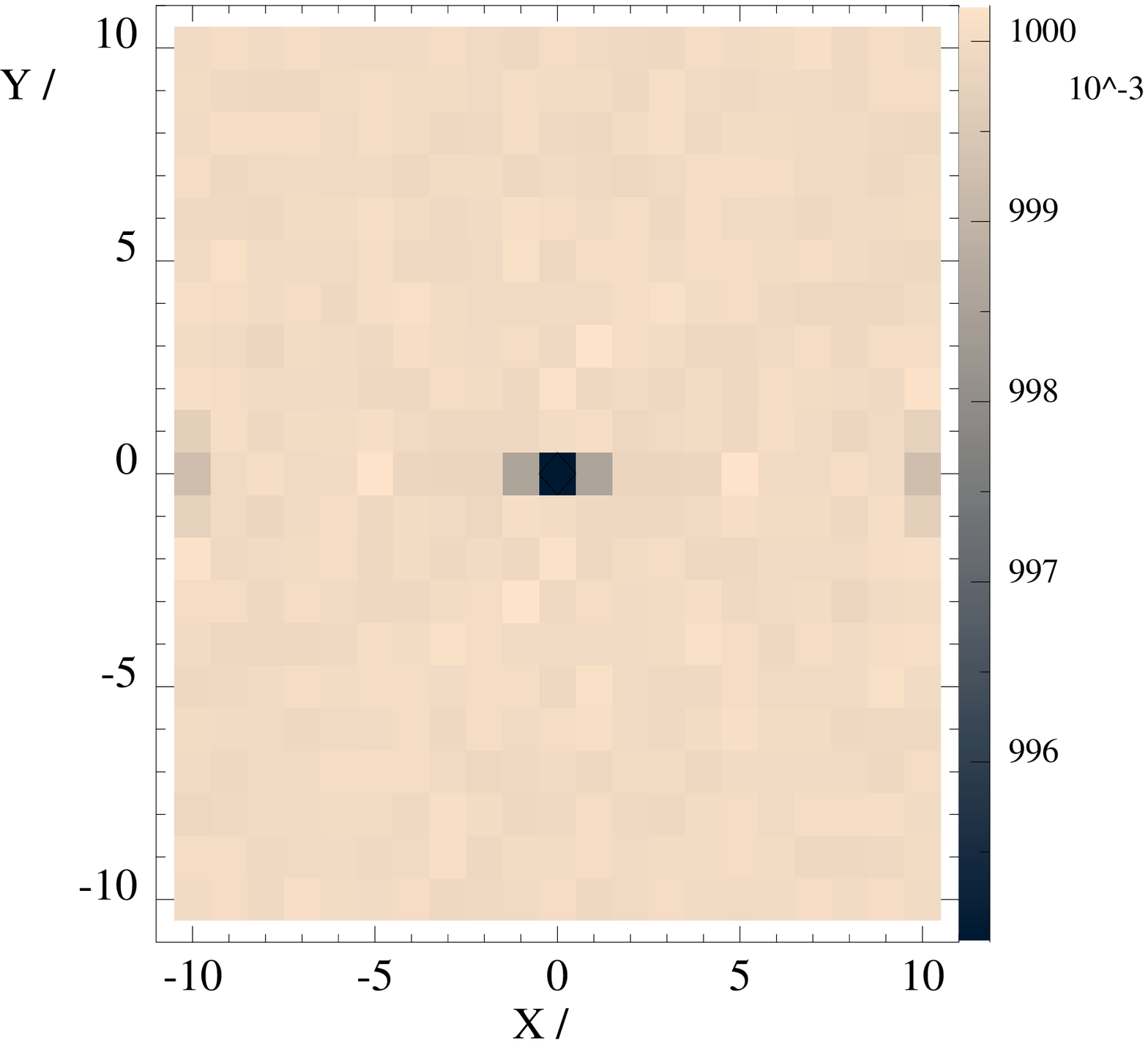}}
%\subfigure[$u_{x,0}=0.1$]{\includegraphics[width=.3\textwidth]{wi_d2q9guuxux.eps}}
\subfigure[$u_{x,0}=0.1$]{\includegraphics[width=.3\textwidth]{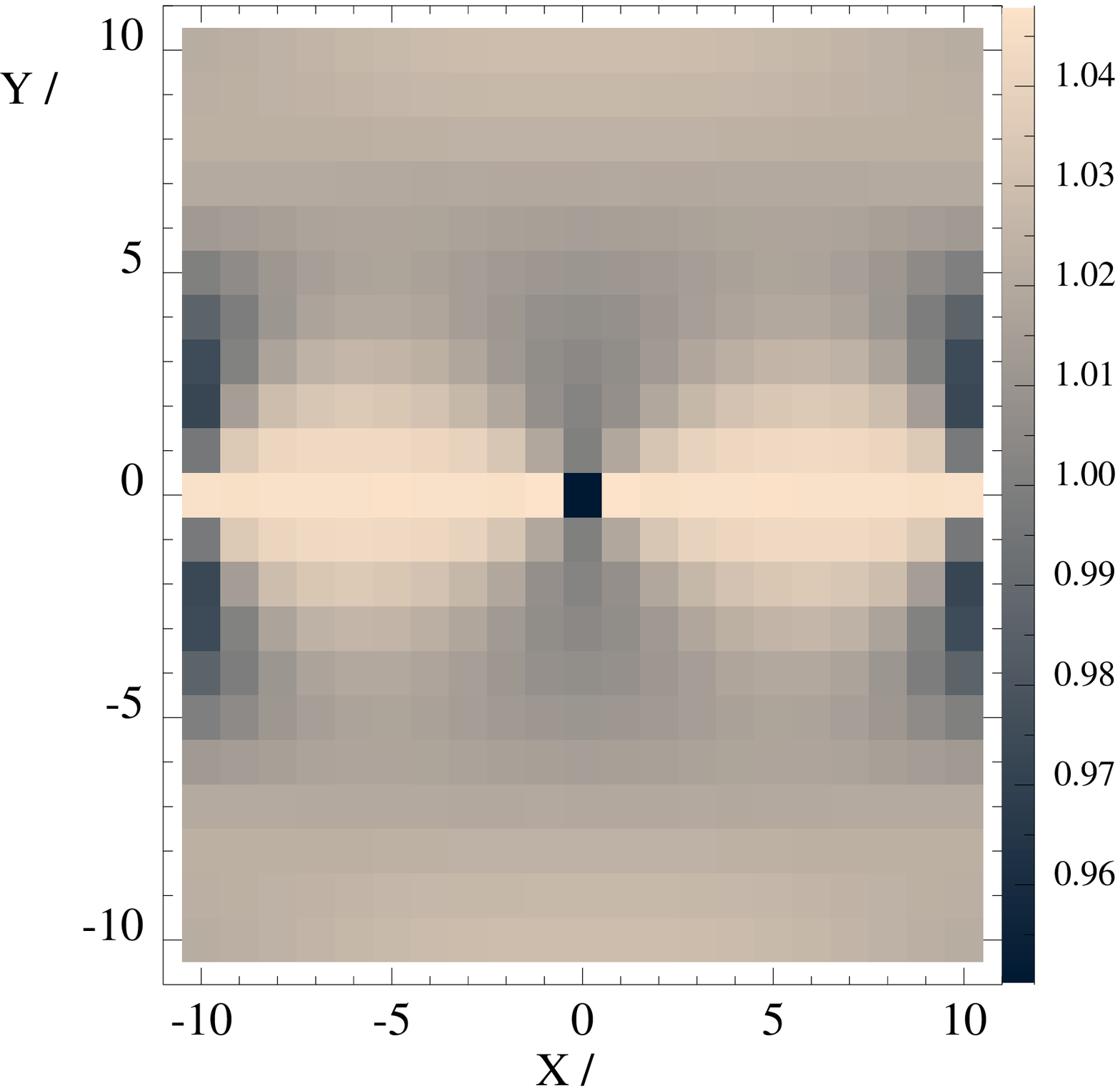}}
\subfigure[$u_{x,0}=0.2$]{\includegraphics[width=.3\textwidth]{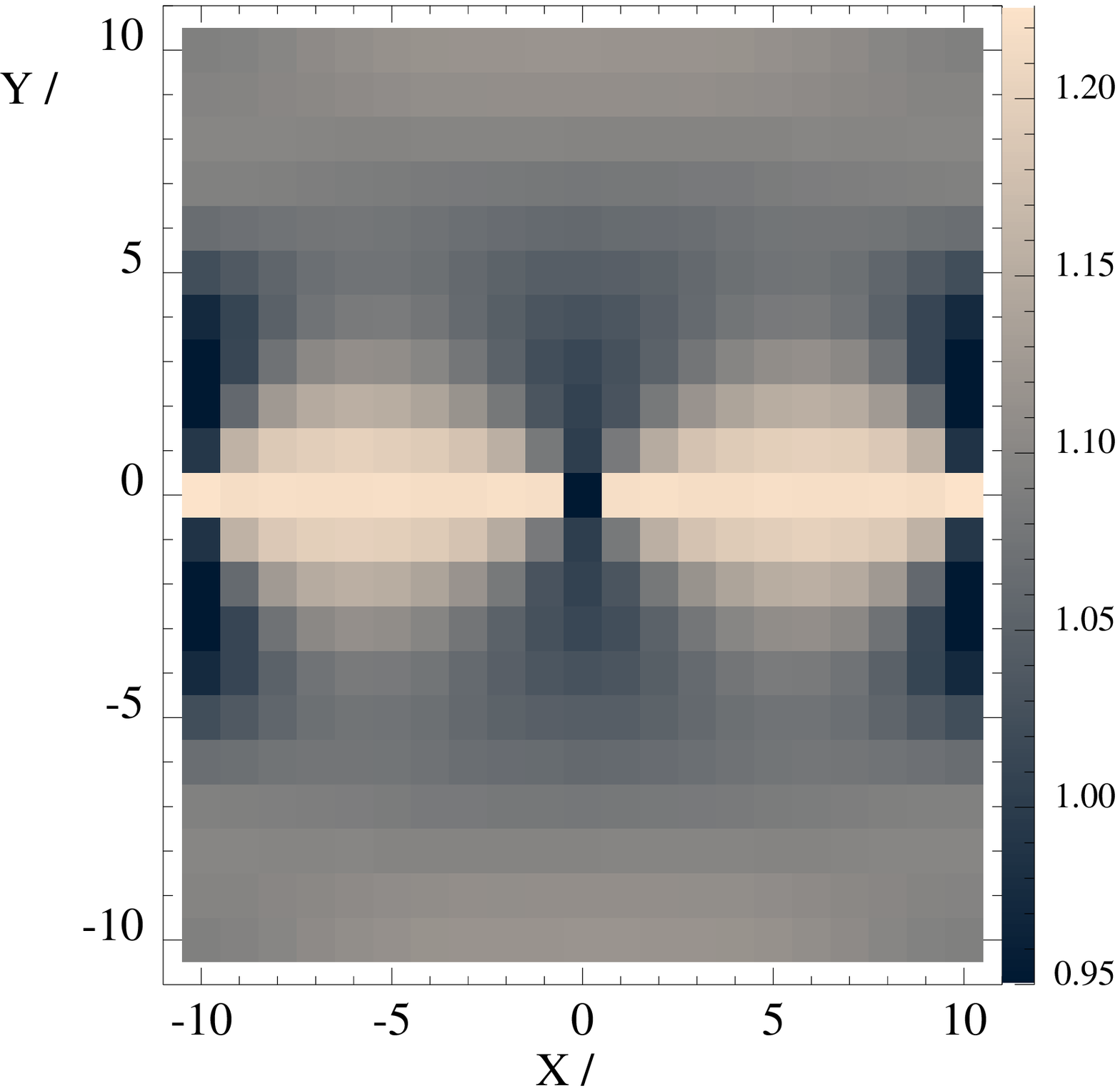}}
\end{psfrags}
\caption{Static structure factor $S_{\vecc{k}}(j_x)$ at different velocities measured for the Hermite norm.}
\label{fig:wsuxux}
\end{figure}

\begin{figure}
\begin{psfrags}
\psfrag{X /   }[Bl][Bc]{$k_x$}
\psfrag{Y /   }[Bl][Bc]{$k_y$}
\psfrag{ 10^-4}[Bl][Bc]{$10^{-4}$}
\subfigure[$u_{x,0}=0.0$]{\includegraphics[width=.3\textwidth]{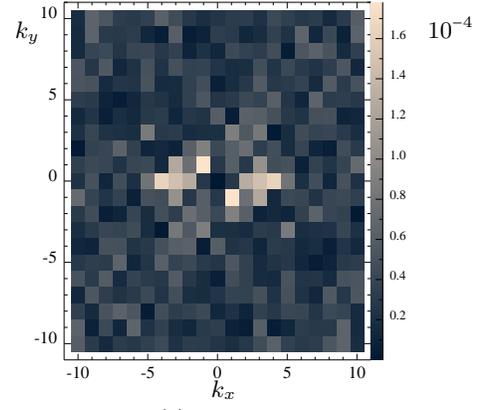}}
%\subfigure[$u_{x,0}=0.1$]{\includegraphics[width=.3\textwidth]{wi_d2q9guuxuy.eps}}
\subfigure[$u_{x,0}=0.1$]{\includegraphics[width=.3\textwidth]{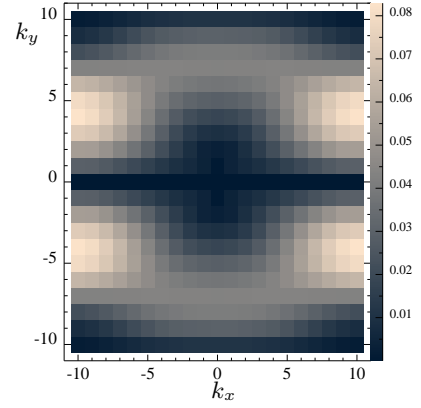}}
\subfigure[$u_{x,0}=0.2$]{\includegraphics[width=.3\textwidth]{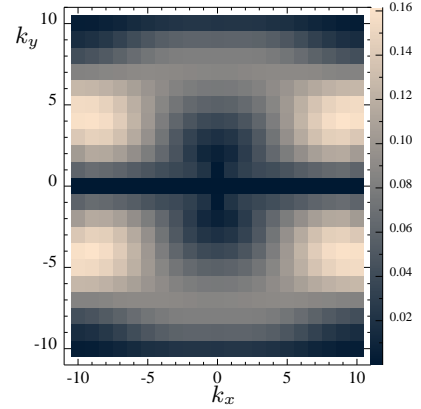}}
\end{psfrags}
\caption{Cross correlator $R_{\vecc{k}}(j_x, j_y)$ at different velocities measured for the Hermite norm.}
\label{fig:wsuxuy}
\end{figure}

In Figs.~\ref{fig:wsrr}, \ref{fig:wsuxux}, and \ref{fig:wsuxuy} we observe that the correlators lose the relatively good agreement with the isotropy requirement of the ideal gas, i.e. the wave number independence as we increase the velocity. 
% These deviations do not approach those of the errors in the original Ladd implementation \cite{ladd-1993} for short length scales \cite{kaehler-2009} that omitted the noise on the ghost modes. However, 
They are sensitive to increased velocities and isotropy at the correlations is destroyed. Errors are not limited to large $\mathbf{k}$ and impinge on the hydrodynamic ($\mathbf{k}$ small) region. Different correlators violate isotropy at different length scales and directions but we can generalize that the violations for certain length scales and spatial directions exceed those observed on the level of the ensemble averaged correlations discussed so far. As an example the density correlator $S_{\vecc{k}}(\rho)$ deviates by more than $20\%$ on all length scales in the $x$ direction  at $u_{x,0} = 0.2$ in Fig.~\figref{fig:wsrr}(c) while the ensemble average finds a deviation of about $6\%$ in Fig.~\figref{fig:wdmadma}. Comparing Figs.~\ref{fig:wsrr}, \ref{fig:wsuxux}, and \ref{fig:wsuxuy} at $u_{x,0} = 0.2$ with $u_{x,0} = 0.1$ we observe that the structure of the anisotropy is largely independent of the average system speed although there are small deviations. Another observation is that although $\langle j_x j_y\rangle$ is small compared to other cross correlators in Fig.~\figref{fig:wdmadmbfit1} this is mostly due to a fortuitious cancellation of errors for different values of $k$. The absolute deviations for the $\langle \delta j_x (k) \delta j_y (k) \rangle$ are of similar magnitude compared to $\langle \delta j_x (k) \delta j_x (k) \rangle$.

%The normalization constant $N^{ab}$ is chosen such that $S_{\vecc{k}}(M^a) = 1$ and $R_{\vecc{k}}(M^a, M^b) = 1$ are equivalent to $S_{\vecc{k}}(M^a) = \rho$ and $R_{\vecc{k}}(M^a, M^b) = \rho$. For the density $N^\rho = \frac{1}{\bar{\rho}^{3} V}$ and velocity components $N^{u_\alpha} = \frac{1}{\bar{\rho} V \theta }$ where $\theta = \frac{1}{3}$ for the isothermal D2Q9 model employed.

In summary we can clearly see that as function of the fluid velocity we observe strong deviations from the identities in ~\eref{eqn:fluctf4} and \eref{eqn:dfdf} and the appearance of off-diagonal correlations which are not present in the case of $\vecc{u} = 0$. We conclude that Galilean invariance is indeed violated and that the fluctuation-dissipation theorem of \eref{eqn:FDT} is not longer diagonalized by the simple choice of $f_i^0 / \rho \approx w_i$ in \eref{eqn:fluctf3}.

%With increasing $u_x$ we observe significant deviations from \eref{fluctf4} not only in amplitude of the cross correlators but we also find correlations between different moments. Surprisingly some of the off-diagonal deviations even exceed the deviations of the cross correlators.

\section{Local Velocity dependent Transforms}

The question now is whether we can alleviate the difficulties we have encountered by avoiding the approximation of $f_i^0(\vecc{u} = 0) = \rho w_i$ in the normalization condition. Removing the velocity dependence in the normalization condition could very likely be the source of the Galilean invariance violations observed. Instead of using \eref{eqn:fluctf5} we now include the velocity dependence of the equilibrium distribution in \eref{eqn:fluctf3}. The orthogonalization condition then becomes
\beq{eqn:fnorm}
\sum_i \tilde{m}_i^a (\vecc{u}) \tilde{m}_i^b(\vecc{u}) w_i \left\lbrack 1 + \frac{1}{\theta} \vecc{u}.v_i + \frac{1}{2\theta^2}\left(\vecc{u}.v_i\right)^2 - \frac{1}{2\theta}\vecc{u}.\vecc{u}\right\rbrack = \delta^{ab}
\eeq
where the velocity $\vecc{u}(\vecc{r}, t)$ is understood to be local to the lattice site $\vecc{r}$. We obtain a new set of transformation matrices $\tilde{m}_i^a$ by starting with the physical moments, $\rho$, $j_x$, $j_y$, $\Pi_{xx-yy}$, $\Pi_{xy}$, $\Pi_{xx+yy}$ and perform a Gram-Schmidt orthogonalization with respect to the new scalar product 
\beq{eqn:fnormsp}
%\langle \vecc{a}, \vecc{b} \rangle = 
\sum_i a_i f_i^0 b_i.
\eeq
The iterative procedure then follows
\beq{eqn:gramschmidt}
\hat{m}_i^a = m_i^a - \sum_{b=0}^{a-1} \tilde{m}_i^b  \sum_j \tilde{m}_j^b f_j^0 m_j^a 
%}{ \sum_k \tilde{m}_k^b f_k^0 \tilde{m}_k^b }
\eeq
with an intermediate normalization step 
\beq{eqn:gsortho}
\tilde{m}_i^a = \frac{\hat{m}_i^a}{\sum_j \hat{m}_j^a f_j^0 \hat{m}_j^a }.
\eeq
With these new matrix elements $\tilde{m}_i^a$ we can define the physically relevant moments
\beq{eqn:Mtilde}
\tilde{M}^a = \sum_i \tilde{m}_i^a f_i.
\eeq

One useful side effect of this transform is that the equilibrium values for all moments other than the density vanish such that
\beq{eqn:fnormmeq}
\tilde{M}^{a,0} = 
\left\{
\begin{array}{cl}
\rho & \text{if } a=0\\
0 & \text{otherwise}
\end{array}
\right.
\eeq
This is a direct consequence of condition \eref{eqn:fnorm} if we recognize that $\tilde{M}^{a,0} = \sum_i \tilde{m}_i^a f_i^0 \tilde{m}_i^0 = \rho \delta^{a0}$ because the density mode is still the one vector $m_i^0 = \tilde{m}_i^0 = 1_i$.
This new process does not alter the hydrodynamic limit of the lattice Boltzmann method because we only alter the moments multiples of $\vecc{u}(\vecc{r})$ with the conserved quantity eigenvectors of the density $1_i$ and momentum modes $v_{i\ga}$. If we interpret the local velocity $\vecc{u}(\vecc{r})$ as an arbitrary constant we do not alter the hydrodynamic equations at all by virtue of our discussion in \cite{kaehler-2011}. We will refer to \eref{eqn:fnorm} simply as the ``$f$-norm'' in the following.

%a Gram-Schmidt orthogonalization to the original multi-relaxation time basis introduced by d'Humieres \cite{dhumieres-1992} given in appendix \ref{sec:appdhumieres}. However, as the new orthogonalization condition we use the full equilibrium distribution of \eref{eqn:f0} 

%where we subtract the projection of the previous moment onto the current moment from the current moment. These projections, however, are moments of the equilibrium distribution ...

%>>>>>> Having 0 for equilibrium moment is identical to having orthogonal basis?<< No, Hermite norm is also orthogonal. What is the extra quality here? Is there a necessity for the equilibrium moments to vanish if we rotate into the $\vecc{u} = 0$ space? Yes, in local transform base the momentum vanishes but what about constant offsets in $epsilon$ and $\Pi_{xx+yy}$? Is this a D2Q9 specific phenomenon or can we construct an argument that makes this generally true? <<<<<<

In order to maintain positive-definiteness of the scalar product \eref{eqn:fnormsp} we must be mindful here of the fact that the normalization constant needs to be positive at all times. The second order expansion of the equilibrium distribution \eref{eqn:f0} we use here, however, is not. For large enough $|\vecc{u}|$ the $f_{i, 0}(\rho, \vecc{u}, \theta) < 0$ and the orthogonalisation has no solution.
\begin{figure}
\includegraphics[width=.3\textwidth]{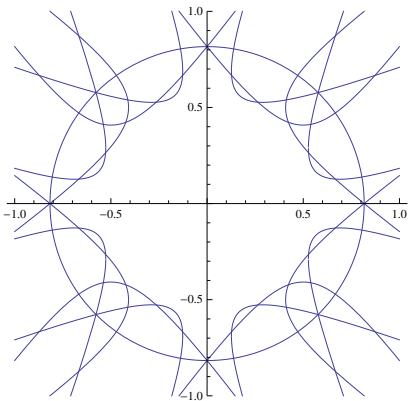}
\caption{$f_i^0(u_{x,0}, u_y)=0$ for all $i$ in the case of the D2Q9 model. In the area inside the curves $f_i > 0$ for all $i$. Outside at least one $f_i < 0$ and consequently the orthogonalization does not find a solution.}
\label{fig:uvalidity}
\end{figure}
In Fig.~\figref{fig:uvalidity} we show the $0$-transition of the second order expansion of the equilibrium distribution in the case of the D2Q9 model as a function of $\vecc{u}$. This plot shows the accessible velocity range. As long as our velocities do not fall outside the central area of Fig.~\figref{fig:uvalidity} the transformation matrix is guaranteed to be positive definite and the Gram-Schmidt will provide a solution.

\begin{figure}
\begin{psfrags} % Optional

\psfrag{XXXXXXXXf0f0lt}[Bl][Br]{\scriptsize\hspace{-0.075\textwidth} $\langle \delta f_0 \delta f_0 \rangle$}
\psfrag{XXXf0f0w}[Bl][Br]{\scriptsize\hspace{-0.075\textwidth} $\langle \delta f_0 \delta f_0 \rangle_H$}
\psfrag{XXXf0f0full}[Bl][Br]{\scriptsize\hspace{-0.075\textwidth} $\langle \delta f_0 \delta f_0 \rangle_f$}

\includegraphics[width=0.3\textwidth, angle = 270]{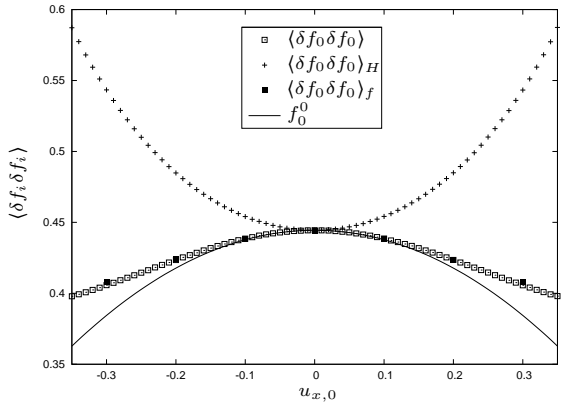}
\end{psfrags}
% Optional
\caption{$\langle \lpar \delta f_0 \rpar^2 \rangle$ in a $21\times21$ D2Q9 fluctuating LB simulation employing the $f$-norm with look up tables. Equilibrium moments are calculated to third order. $\langle \delta f_0 \delta f_0 \rangle_f$ are datapoints taken from a fully local implementation that foregoes the look up table solution. We plot the equilibrium distribution $f_0^0$ and the Hermite norm correlator $\langle \delta f_0 \delta f_0 \rangle_H$ for comparison.}
\label{fig:fdf0df0}
\end{figure}

\begin{figure}
\begin{psfrags} % Optional

\psfrag{XXXf1f1}[Bl][Br]{\scriptsize\hspace{-0.075\textwidth} $\langle \delta f_1 \delta f_1 \rangle$}
\psfrag{XXXf1f1full}[Bl][Br]{\scriptsize\hspace{-0.075\textwidth} $\langle \delta f_1 \delta f_1 \rangle_f$}
\psfrag{XXXf2f2}[Bl][Br]{\scriptsize\hspace{-0.075\textwidth} $\langle \delta f_2 \delta f_2 \rangle$}
\psfrag{XXXf3f3full}[Bl][Br]{\scriptsize\hspace{-0.075\textwidth} $\langle \delta f_3 \delta f_3 \rangle_f$}
\psfrag{XXXf3f3}[Bl][Br]{\scriptsize\hspace{-0.075\textwidth} $\langle \delta f_3 \delta f_3 \rangle$}
\psfrag{XXXf2f2f4f4full}[Bl][Br]{\scriptsize\hspace{-0.075\textwidth} $\langle \delta f_2 \delta f_2 \rangle_f$}

\includegraphics[width=0.3\textwidth, angle = 270]{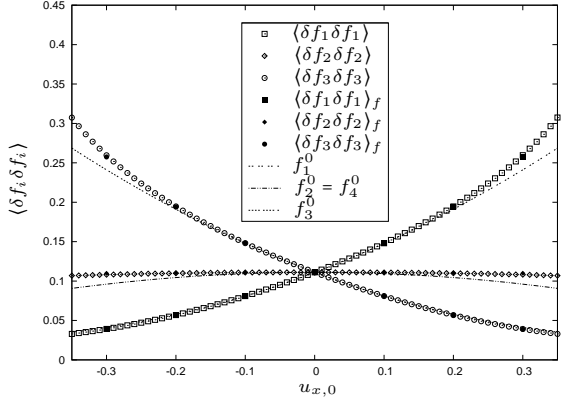}
\end{psfrags}
% Optional
\caption{$\langle  \lpar \delta f_i \rpar^2 \rangle$ for $i=1...3$ in a $21\times21$ D2Q9 fluctuating LB simulation employing the $f$-norm. We plot $f_i^0$ for comparison. $\langle  \lpar \delta f_4 \rpar^2 \rangle$ is not shown as it appears identical to $\langle  \lpar \delta f_2 \rpar^2 \rangle$ within the scale of this plot.}
\label{fig:fdf14df14}
\end{figure}

\begin{figure}
\begin{psfrags} % Optional
\includegraphics[width=0.3\textwidth, angle = 270]{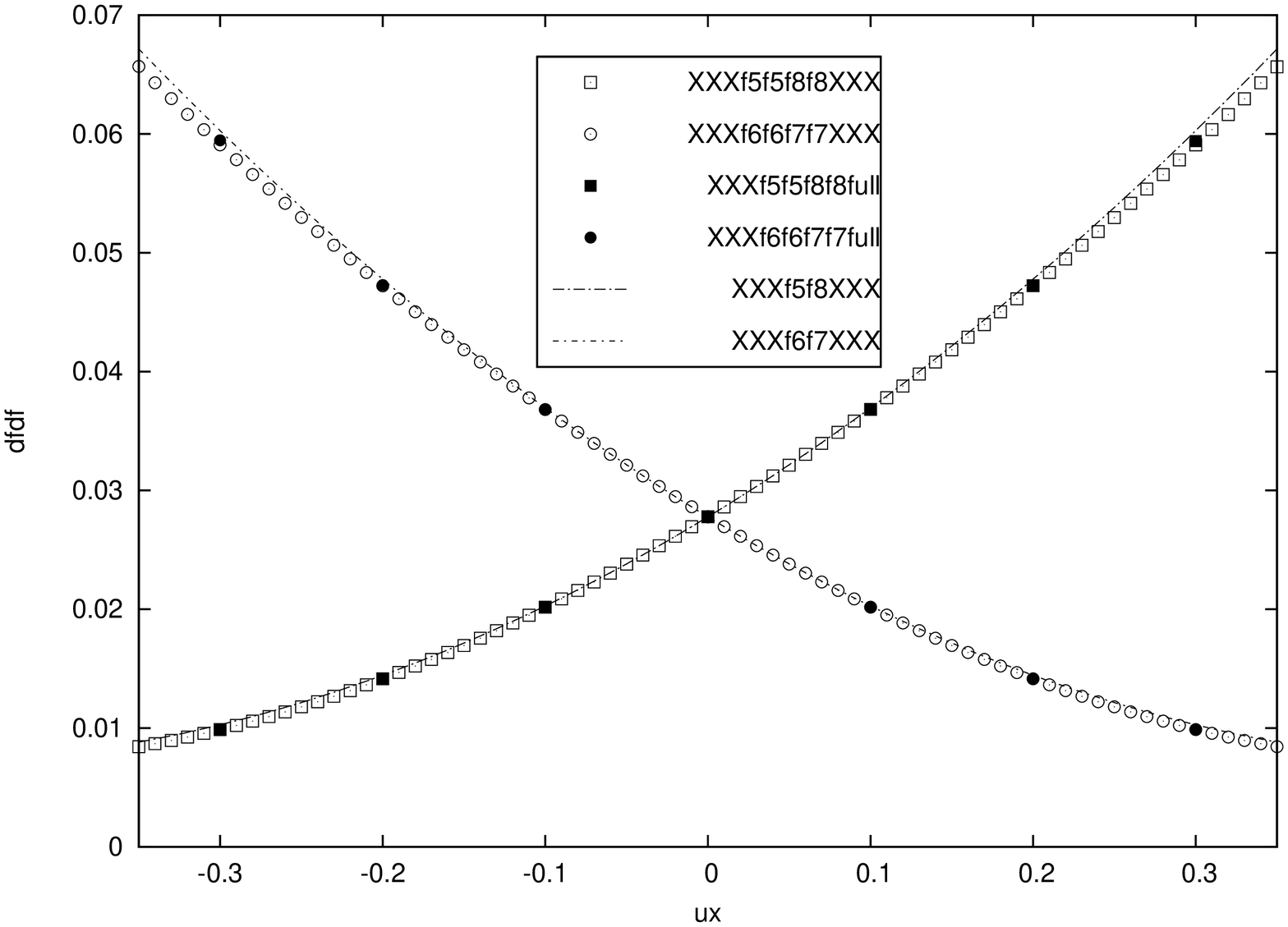}
\end{psfrags}
% Optional
\caption{$\langle \lpar \delta f_i \rpar^2 \rangle$ for $i=5...8$ in a $21\times21$ D2Q9 fluctuating LB simulation employing the $f$-norm. We plot $f_i^0$ for comparison. $\langle \lpar \delta f_8 \rpar^2 \rangle$ and $\langle \lpar \delta f_7 \rpar^2 \rangle$ are not shown as they appears identical to $\langle \lpar \delta f_5 \rpar^2 \rangle$ and $\langle \lpar \delta f_6 \rpar^2 \rangle$ respectively in the scale of this plot.}
\label{fig:fdf58df58}
\end{figure}

The matrix elements $\tilde{m}_i^a(\vecc{u}(\vecc{r}))$ we obtain are now functions of the local velocity $\vecc{u}(\vecc{r})$ at lattice site $\vecc{r} = \lpar x, y \rpar^T$. In principle they have to be evaluated at every lattice site during every update cycle. We have implemented a fluctuating LB simulation with these matrices and the results are encouraging in that Galilean invariance violations are significantly smaller. Some results of these are shown in Figs.~\figref{fig:fdf0df0}, \figref{fig:fdf14df14}, and \figref{fig:fdf58df58}. However, even in the relatively simple D2Q9 model the matrix elements of higher order moments are polynomials of $O(\vecc{u}^{16})$ and therefore the local evaluation of these matrix elements becomes prohibitively costly. Our test implementation used between $95\%$ and $99\%$ of the computation time of an update cycle in the evaluation of the local transforms.

One might think that going to the full second order expansion of $f_i^0$ might not be necessary and going only to first order in $\mathbf{u}$ would make the structure of the matrix elements significantly simpler. However, working with only the first order expansion introduces anisotropy effects between the different spatial axis. Removing these effectively makes the expressions for the $\tilde{m}_i^a$ even more complicated than the regular second order expressions where our Gram-Schmidt orthogonalization renders the moments isotropic.

It is, however, not strictly necessary to calculate the transforms to machine precision. Judging from our observations of the Hermite norm implementation it is sufficient to calculate tables of the matrix elements on a velocity grid with velocities $\vecc{u}_g(g_\ga)$ where $g_\ga$ is the grid position and use these matrix elements from a look up table in the transforms. The benefit is practicality, the pay off is that we may not quite obtain the same amount of improvement we might expect to find otherwise. 
One caveat is that we lose the convenient form of the equilibrium moments in \eref{eqn:fnormmeq}. In fact the projection of the moments in the representation of current local velocity to that of the nearest look up table velocity becomes algebraically similarly complex as the calculation of the matrix elements themselves. However, as we are concerned with a second order theory here we choose to only use terms of up to $O\lpar\vecc{u}_g^3\rpar$. While we do not change the conserved quantities we do change the stress and ghost moments at orders $O\lpar\vecc{u}^4\rpar$ and higher and thus introduce small errors. An example of these equilibrium moments and the matrix transform elements for D2Q9 can be found in \cite{kaehler-2012-notebook}.

The velocity grid spacing for the look up table can be relatively coarse. It is helpful if the entire look up table of velocities can fit into the second level cache of the CPU the simulation is run on. In our D2Q9 test case we typically use a $51 \times 51$ grid with $-0.5 \le u_{g,x} \le 0.5$, $-0.5 \le u_{g,y} \le 0.5$, and $\Delta u_g = 0.02$. Comparing this velocity range with Fig.~\figref{fig:uvalidity} we notice that the corners of this square in velocity space falls outside the valid $f_i^0(\mathbf{u}) > 0$ range. The matrix elements here are simply evaluated to ``not a number'' and the simulation fails once any one of these velocities are reached. In principle one could also catch outliers in the velocity and just choose the matrix elements for a smaller velocity. The moment projection would still function. However, this would alter the algorithm and the results would not be reliable representations of the method discussed here. For applications, especially at high velocities and low densities it will be necessary to include such an exception handling routine.

One could argue that we might as well have just calculated the matrix elements to a lower order directly, forego the matrix element look up tables and use the original simple equilibrium moments. However, in that case we would violate conservation laws and the calculation of the $2q^2$ matrix element polynomials is still significantly more expensive than the evaluation of $q-d-1$ non-conserved moments in a D$d$Q$q$ lattice Boltzmann configuration.

\begin{figure}
\begin{psfrags} % Optional
\includegraphics[width=0.3\textwidth, angle = 270]{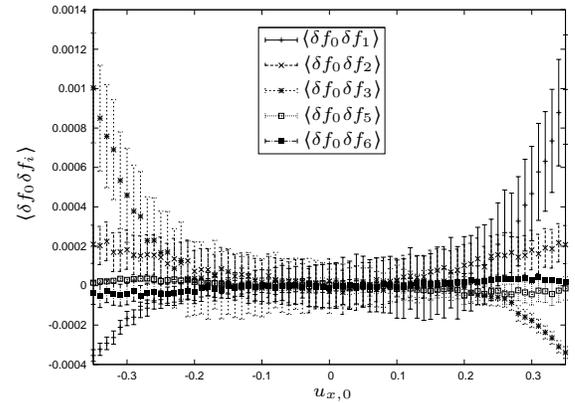}
\end{psfrags}
% Optional
\caption{$\langle \delta f_0 \delta f_i \rangle$ for $i=1...8$ in a $21\times21$ D2Q9 fluctuating LB simulation employing the $f$-norm. }
\label{fig:fdf0dfi}
\end{figure}

To evaluate the implementation of the $f$-norm we perform the same measurements we did for the Hermite norm. We use a D$2$Q$9$ ideal gas simulation with periodic boundaries, and a side length of 21. In Fig.~\figref{fig:fdf0df0} we observe the same $\langle  \delta f_0 \delta f_0  \rangle$ correlator we did in Fig.~\figref{fig:wdf0df0}. We find that with the $f$-norm the trend actually does follow the $f_0^0$ prediction and within $-0.2 \le u_{x,0} \le 0.2$ we are in good agreement with $f_0^0$ but at larger speeds we find smaller but noticeable deviations. In Figs.~\ref{fig:fdf14df14},~\ref{fig:fdf58df58} we find much better agreement for all other distribution function correlation functions for the $f$-norm compared to the Hermite norm in Figs.~\ref{fig:wdf14df14}, and~\ref{fig:wdf58df58}. Again we notice very good agreement for $|u_x| \le 0.2$. 

The remaining deviations from the equilibrium distributions we find with the $f$-norm are not an artifact of either the look up table method or the third order expansion of the equilibrium moments. We performed the same measurement with the fully locally orthogonalized set of transforms, albeit with fewer data points due to the much higher computational effort involved. $\langle \delta f_i \delta f_i \rangle_f$ in Figs.~\figref{fig:fdf0df0},~\figref{fig:fdf14df14}, and \figref{fig:fdf58df58} indicate that the deviations from the equilibrium distributions can indeed not be explained with either the look up table method or the cut off on the equilibrium moments as the results obtained form the look up table method with third order equilibrium moments appears to be consistent from the fully locally orthogonalized $f$-norm.

\begin{figure}
\begin{psfrags} % Optional
\psfrag{XXXXXXXXXXXXXXXXXRR}[Bl][Br]{\scriptsize\hspace{-0.1\textwidth} $\langle \delta \tilde{\rho} \delta\tilde{\rho} \rangle$}
\psfrag{XXXJXJX}[Bl][Br]{\scriptsize\hspace{-0.10\textwidth} $\langle \delta \tilde{j}_x \delta \tilde{j}_x \rangle$}
\psfrag{XXXJYJY}[Bl][Br]{\scriptsize\hspace{-0.10\textwidth} $\langle \delta \tilde{j}_y \delta \tilde{j}_y \rangle$}
\psfrag{XXXPXXMPYY}[Bl][Br]{\scriptsize\hspace{-0.1\textwidth} $\langle \lpar \delta \tilde{\Pi}_{xx-yy} \rpar ^2 \rangle$}
\psfrag{XXXPXY}[Bl][Br]{\scriptsize\hspace{-0.10\textwidth} $\langle \lpar \delta \tilde{\Pi}_{xy} \rpar^2 \rangle$}
\psfrag{XXXPXXPPYY}[Bl][Br]{\scriptsize\hspace{-0.10\textwidth} $\langle \lpar \delta \tilde{\Pi}_{xx+yy} \rpar^2 \rangle$}
\psfrag{XXXG1G1}[Bl][Br]{\scriptsize\hspace{-0.10\textwidth} $\langle \delta \tilde{q}_x \delta \tilde{q}_x \rangle$}
\psfrag{XXXG2G2}[Bl][Br]{\scriptsize\hspace{-0.10\textwidth} $\langle \delta \tilde{q}_y \delta \tilde{q}_y \rangle$}
\psfrag{XXXG3G3}[Bl][Br]{\scriptsize\hspace{-0.10\textwidth} $\langle \delta \tilde{\epsilon} \delta \tilde{\epsilon} \rangle$}
\includegraphics[width=0.3\textwidth, angle = 270]{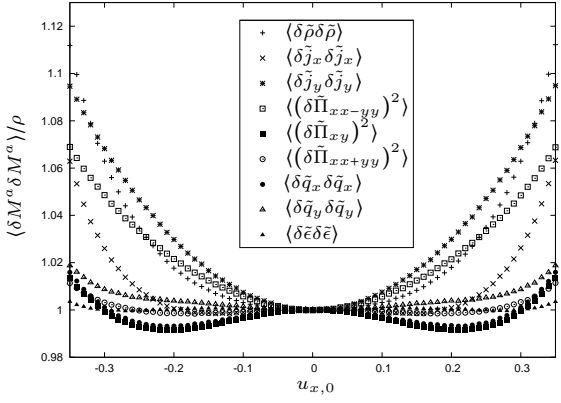}
\end{psfrags}
% Optional
\caption{Correlators $\langle \delta \tilde{M}^a \delta \tilde{M}^a\rangle_f$ normalized to $\rho$ according to \eref{eqn:fluctf4} in a $21\times21$ D2Q9 fluctuating LB simulation employing the $f$-norm.}
\label{fig:fdmadma}
\end{figure}

\begin{figure}
\begin{psfrags} % Optional
\psfrag{XXXXXXXXXXXXXXXXXRR}[Bl][Br]{\scriptsize\hspace{-0.1\textwidth} $\langle \delta \tilde{\rho} \delta\tilde{\rho} \rangle$}
\psfrag{XXXJXJX}[Bl][Br]{\scriptsize\hspace{-0.10\textwidth} $\langle \delta \tilde{j}_x \delta \tilde{j}_x \rangle$}
\psfrag{XXXJYJY}[Bl][Br]{\scriptsize\hspace{-0.10\textwidth} $\langle \delta \tilde{j}_y \delta \tilde{j}_y \rangle$}
\psfrag{XXXPXXMPYY}[Bl][Br]{\scriptsize\hspace{-0.1\textwidth} $\langle \lpar \delta \tilde{\Pi}_{xx-yy} \rpar ^2 \rangle$}
\psfrag{XXXPXY}[Bl][Br]{\scriptsize\hspace{-0.10\textwidth} $\langle \lpar \delta \tilde{\Pi}_{xy} \rpar^2 \rangle$}
\psfrag{XXXPXXPPYY}[Bl][Br]{\scriptsize\hspace{-0.10\textwidth} $\langle \lpar \delta \tilde{\Pi}_{xx+yy} \rpar^2 \rangle$}
\psfrag{XXXG1G1}[Bl][Br]{\scriptsize\hspace{-0.10\textwidth} $\langle \delta \tilde{q}_x \delta \tilde{q}_x \rangle$}
\psfrag{XXXG2G2}[Bl][Br]{\scriptsize\hspace{-0.10\textwidth} $\langle \delta \tilde{q}_y \delta \tilde{q}_y \rangle$}
\psfrag{XXXG3G3}[Bl][Br]{\scriptsize\hspace{-0.10\textwidth} $\langle \delta \tilde{\epsilon} \delta \tilde{\epsilon} \rangle$}
\includegraphics[width=0.3\textwidth, angle = 270]{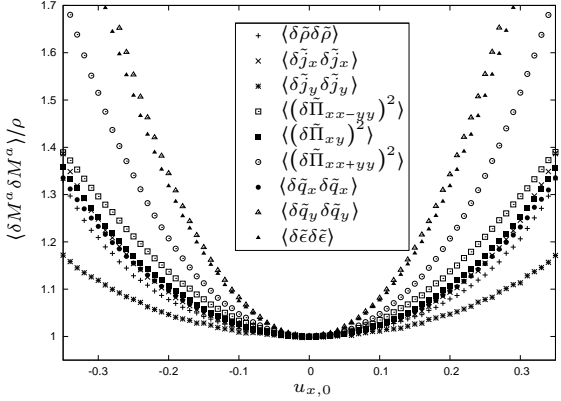}
\end{psfrags}
% Optional
\caption{Correlators $\langle \delta \tilde{M}^a \delta \tilde{M}^a\rangle$ normalized to $\rho$ according to \eref{eqn:fluctf4} measured in a $21\times21$ D2Q9 fluctuating LB simulation employing the Hermite norm.}
\label{fig:wfdmadma}
\end{figure}

%\begin{figure}
%\label{fig:fdmadm0}
%\begin{psfrags} % Optional
%\includegraphics[width=0.3\textwidth, angle = 270]{fi_m0ma.eps}
%\end{psfrags}
% Optional
%\caption{Correlators $\langle \delta M^0 \delta M^b\rangle$ for $a=1...8$ normalized to $\rho$ according to% \eref{eqn:fluctf4} in a $21\times21$ D2Q9 fluctuating LB simulation employing the $f$-norm.}
%\end{figure}

Measuring the moment space correlators in the $f$-norm poses an interesting question. Do we measure with respect to the Hermite norm or the $f$-norm and in the case of the latter with respect to which velocity? To answer this question we conduct a thought experiment. $\delta M^a$ should be Galilean invariant for any $a$, in particular the momentum components. In the Hermite norm we have
\beq{eqn:widjx}
\delta j_{x} = \sum_i m_i^a f_i - \sum_i m_i^a f_i^0 = \sqrt{3} \lpar \rho u_x - \rho_0 u_{x,0}\rpar
\eeq
and for the $f$-norm
\beq{eqn:fidjx}
\delta \tilde{j}_{x} = \sum_i \tilde{m}_i^a f_i - \sum_i \tilde{m}_i^a f_i^0 = \sqrt{3} \rho \lpar u_x - u_{x,0}\rpar.
\eeq
Again $\mathbf{u}_0$ is the mean velocity in the system and $\mathbf{u}$ the local velocity at a given lattice site. If we set $\mathbf{u}_0 = 0$ we have $\delta j_{x} = \delta \tilde{j}_{x} = \sqrt{3}\rho u_x$. Introducing a constant velocity offset $-\mathbf{u}_O$ should leave $\delta j_x$ Galilean invariant, i.e. we expect $\mathbf{u} \rightarrow \mathbf{u} - \mathbf{u}_O$. If we now interpret $\mathbf{u}_0$ as such an offset the Hermite norm is clearly not Galilean invariant under velocity offsets as it introduces an extra $ u_{x,0} \lpar \rho_0 - \rho \rpar$ in \eref{eqn:widjx} whereas the $f$-norm in \eref{eqn:fidjx} behaves as required. Consequently we use the $f$-norm as it provides the correct measurements that leave the $\delta \tilde{M}^a$ invariant under Galilean transformations. Furthermore we measure with respect to the average system velocity $\mathbf{u}_0$ and average density $\rho_0$. Measuring with respect to the local velocity $\mathbf{u}$ and density $\rho$ is nonsensical as $\delta \rho = 0$ and $\delta \mathbf{j} = 0$ in this case. We thus use the $f$-norm such that $\tilde{m}_i^a \tilde{m}_i^b \langle f_i \rangle = \delta^{ab}$ where we make the approximation of \eref{eqn:efif0} $\langle f_i \rangle = f_i^0(\rho_0, \mathbf{u_0})$.

%When measuring the moment correlators one needs to pay close attention which transforms to use to obtain the moment correlators. The Hermite norm, the $f$-norm and if the latter with respect to which velocity? The Hermite norm is clearly inappropriate and the local $f$-norm is not applicable since $\delta M$ would vanish for all moments according to \eref{eqn:fnormmeq}. Instead because of \eref{eqn:dfdef} we should use a norm which is orthogonal with respect to $m_i^a m_i^b \langle f_i \rangle$. Here, as in \eref{eqn:efif0}, we make the approximation $\langle f_i \rangle = f_i^0(\rho_0, \mathbf{u_0})$. Thus, in order to perform moment space measurements in the $f$-norm, we use the transforms evaluated at $f_i^0(\rho_0, \mathbf{u_0})$.

Much like the distribution function correlators the moment correlators $\langle (\delta M^a)^2 \rangle$ shown in Fig.~\figref{fig:fdmadma} exhibit significant improvement compared to those of the Hermite norm  in Fig.~\figref{fig:wdmadma}. This improvement is smaller than the general trend of the distribution function correlators would imply for some modes. In particular the $\langle (\delta \tilde{\rho})^2 \rangle $, $\langle (\delta \tilde{\Pi}_{xx-yy})^2 \rangle $, and $\langle (\delta \tilde{j}_y)^2 \rangle $ correlators deviate significantly for larger $u_x$. Their overall decrease is about $1/3$ compared to the Hermite norm. To make a valid comparison between moment correlators computed in the $f$-norm and the Hermite norm one needs to ensure that for both measurements the moments are obtained in the same way. We therefore measure the moments obtained in a Hermite norm simulation with the $f$-norm evaluated at $\mathbf{u}_0$ in Fig.~\figref{fig:wfdmadma}. We observe that for all moments but $\langle \delta \tilde{\rho} \delta \tilde{\rho} \rangle$ and $\langle \delta \tilde{j}_y \delta \tilde{j}_y \rangle$ the deviations are larger than those measured in the Hermite norm.

\begin{figure}
%\subfigure[Linear coefficient $l$, Hermite norm]{
%\begin{psfrags} % Optional
%\includegraphics[width=0.3\textwidth, angle = 0]{wi_m_matrix1.eps}
%\end{psfrags}
%}
\subfigure[Linear coefficient $l$, $f$-norm]{
\begin{psfrags} % Optional
\psfrag{X /   }[Br][Bc]{}
\psfrag{Y /   }[Br][Bc]{}
\psfrag{jx}[Bc][Bc]{\scriptsize$\tilde{j}_x$\normalsize}
\psfrag{qx}[Bc][Bc]{\scriptsize$\tilde{q}_x$\normalsize}
\psfrag{pxy}[Bc][Bc]{\scriptsize$\tilde{\Pi}_{\times}$\normalsize}
\psfrag{rho}[Bc][Bc]{\scriptsize$\tilde{\rho}$\normalsize}
\psfrag{ppl}[Bc][Bc]{\scriptsize$\tilde{\Pi}_{+}$\normalsize}
\psfrag{pmn}[Bc][Bc]{\scriptsize$\tilde{\Pi}_{-}$\normalsize}
\psfrag{eps}[Bc][Bc]{\scriptsize$\tilde{\epsilon}$\normalsize}
\psfrag{jy}[Bc][Bc]{\scriptsize$\tilde{j}_y$\normalsize}
\psfrag{qy}[Bc][Bc]{\scriptsize$\tilde{q}_y$\normalsize}
\includegraphics[width=0.3\textwidth, angle = 0]{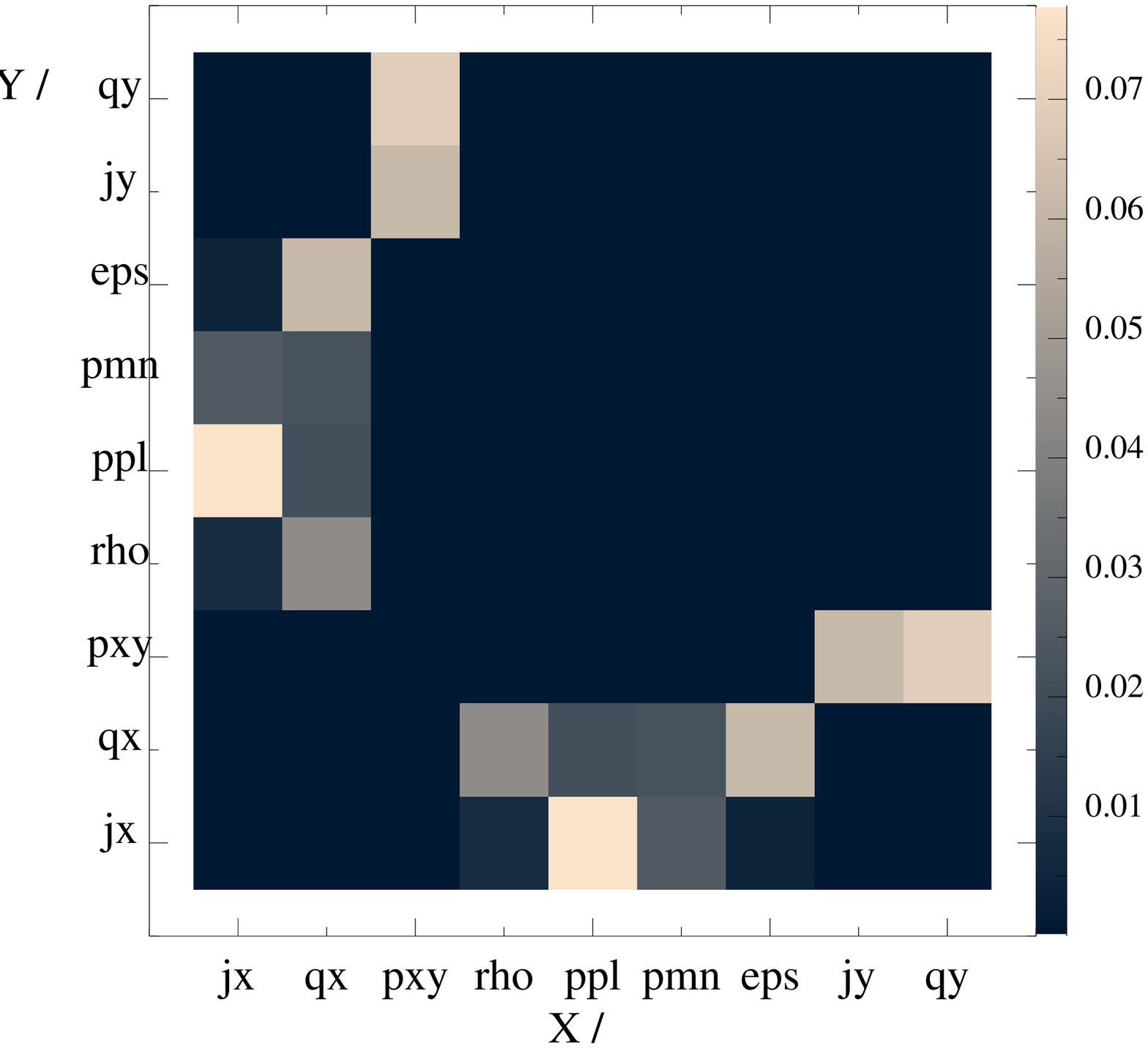}
\end{psfrags}
}
%\subfigure[Quadratic coefficient $q$, Hermite norm]{
%\begin{psfrags} % Optional
%\includegraphics[width=0.3\textwidth, angle = 0]{wi_m_matrix2.eps}
%\end{psfrags}
%}
\subfigure[Quadratic coefficient $q$, $f$-norm]{
\begin{psfrags} % Optional
\psfrag{X /   }[Br][Bc]{}
\psfrag{Y /   }[Br][Bc]{}
\psfrag{jx}[Bc][Bc]{\scriptsize$\tilde{j}_x$\normalsize}
\psfrag{qx}[Bc][Bc]{\scriptsize$\tilde{q}_x$\normalsize}
\psfrag{pxy}[Bc][Bc]{\scriptsize$\tilde{\Pi}_{\times}$\normalsize}
\psfrag{rho}[Bc][Bc]{\scriptsize$\tilde{\rho}$\normalsize}
\psfrag{ppl}[Bc][Bc]{\scriptsize$\tilde{\Pi}_{+}$\normalsize}
\psfrag{pmn}[Bc][Bc]{\scriptsize$\tilde{\Pi}_{-}$\normalsize}
\psfrag{eps}[Bc][Bc]{\scriptsize$\tilde{\epsilon}$\normalsize}
\psfrag{jy}[Bc][Bc]{\scriptsize$\tilde{j}_y$\normalsize}
\psfrag{qy}[Bc][Bc]{\scriptsize$\tilde{q}_y$\normalsize}
\includegraphics[width=0.3\textwidth, angle = 0]{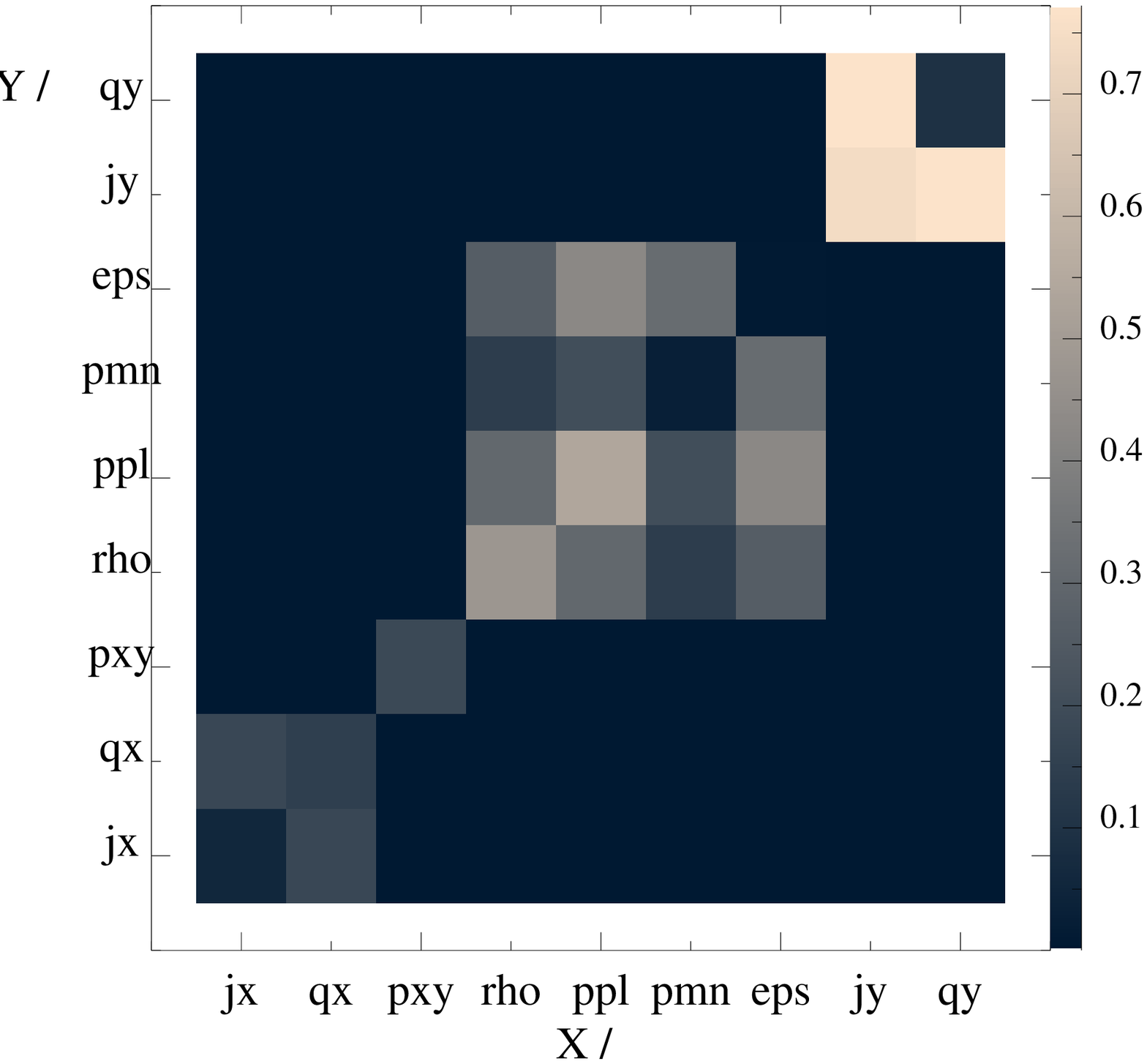}
\end{psfrags}
}
% Optional
%for (a) and (c) and the $f$-norm with look up tables,  $u_g = 0.02$, for (b) and (d). Fit range used was $-0.25 <= u_x <= 0.25 $. We observe a significant decrease in both the linear and quadratic coefficients when employing the $f$-norm.
%}
\caption{Linear and quadratic coefficient $l$ and $q$ of all $81$ ($45$ unique) correlators as a result of fitting $\langle \delta \tilde{M}^a \delta \tilde{M}^b\rangle(u_{x,0})-\delta^{ab}$ to $l u_{x,0} + q u_{x,0}^2$. Brighter color indicates larger coefficients. Moments were reordered to visually identify correlations better. To accommodate for symbol size the stress moments were simplified: $\tilde{\Pi}_\times = \tilde{\Pi}_{xy}, \tilde{\Pi}_- = \tilde{\Pi}_{xx-yy}, \tilde{\Pi}_+ = \tilde{\Pi}_{xx+yy}$). The coefficient at position (0, 1) in image (a) would correspond to linear portion of the $\langle \delta \tilde{j}_x \delta \tilde{q}_x \rangle$ correlator. 
Coefficients were measured on a $21 \times 21$ D2Q9 simulation employing the $f$-norm with look up tables,  $u_g = 0.02$. Fit range used was $-0.25 <= u_x <= 0.25 $.}
\label{fig:fdmadmbfit1}
\end{figure}

\begin{figure}[ht]
\begin{psfrags}
\psfrag{X /   }[Bl][Bc]{$k_x$}
\psfrag{Y /   }[Bl][Bc]{$k_y$}
%\subfigure[$u_x=0.0$]{\includegraphics[width=.3\textwidth]{fi_rr_u00.eps}}
%\subfigure[$u_x=0.1$]{\includegraphics[width=.3\textwidth]{wi_d2q9gurr.eps}}
\subfigure[$u_x=0.1$]{\includegraphics[width=.3\textwidth]{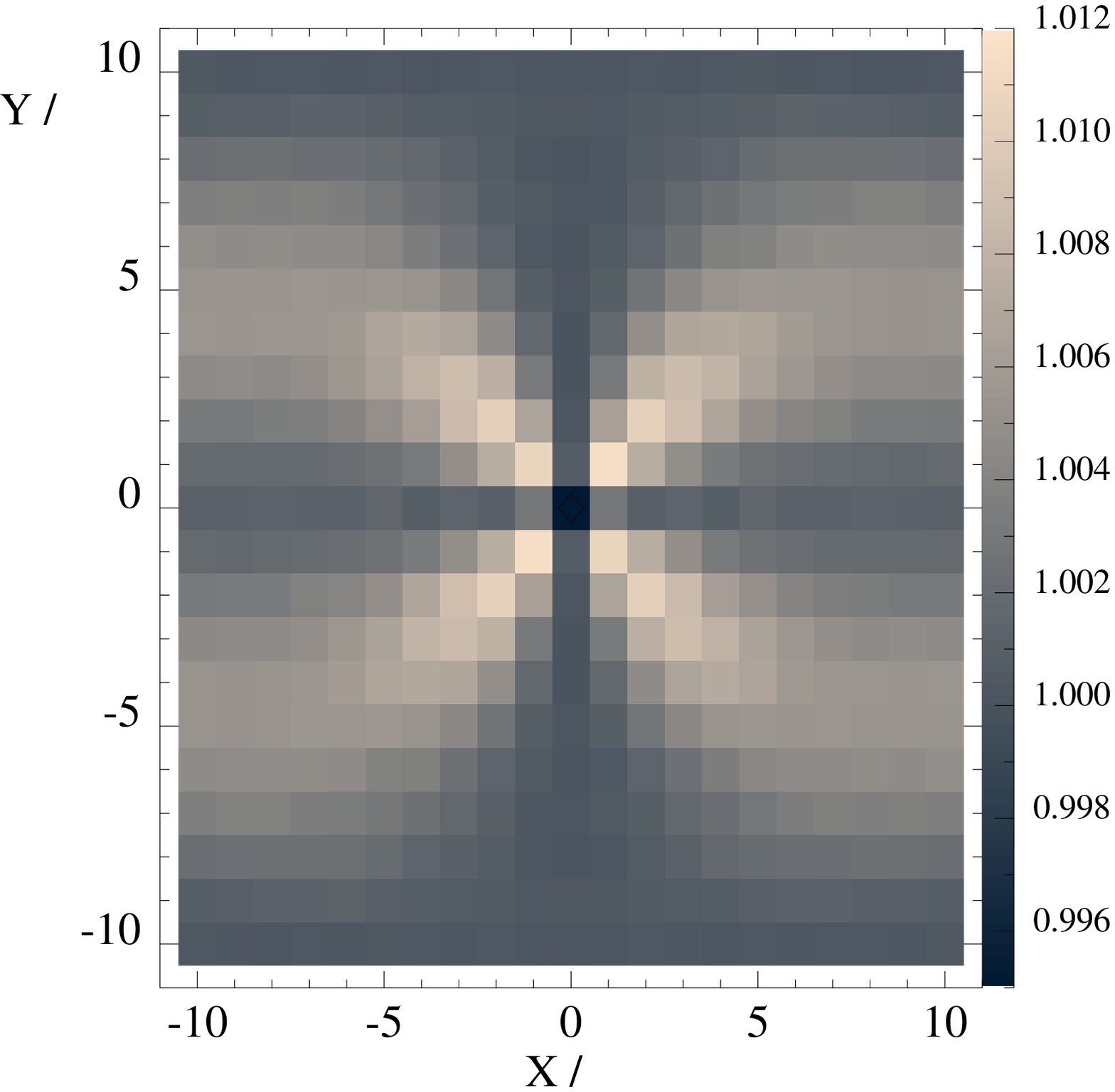}}
\subfigure[$u_x=0.2$]{\includegraphics[width=.3\textwidth]{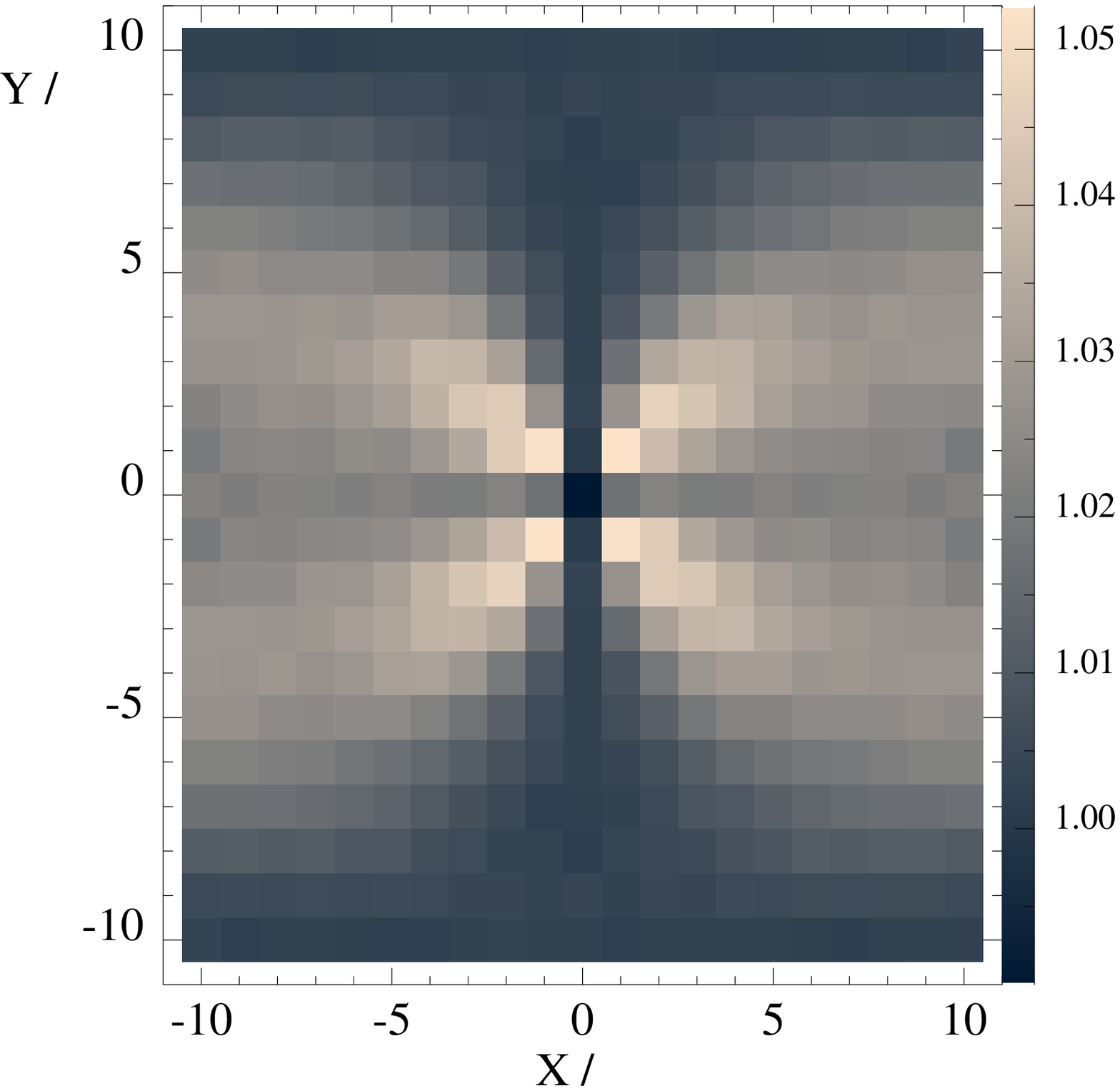}}
\end{psfrags}
\caption{Static structure factor $S_{\vecc{k}}(\tilde{\rho})$ at different velocities measured for the $f$-norm with the look up table and $\Delta u_g = 0.02$.}
\label{fig:fsrr}
\end{figure}

\begin{figure}[ht]
\begin{psfrags}
\psfrag{X /   }[Bl][Bc]{$k_x$}
\psfrag{Y /   }[Bl][Bc]{$k_y$}
%\subfigure[$u_x=0.0$]{\includegraphics[width=.3\textwidth]{fi_uxux_u00.eps}}
%\subfigure[$u_x=0.1$]{\includegraphics[width=.3\textwidth]{wi_d2q9guuxux.eps}}
\subfigure[$u_x=0.1$]{\includegraphics[width=.3\textwidth]{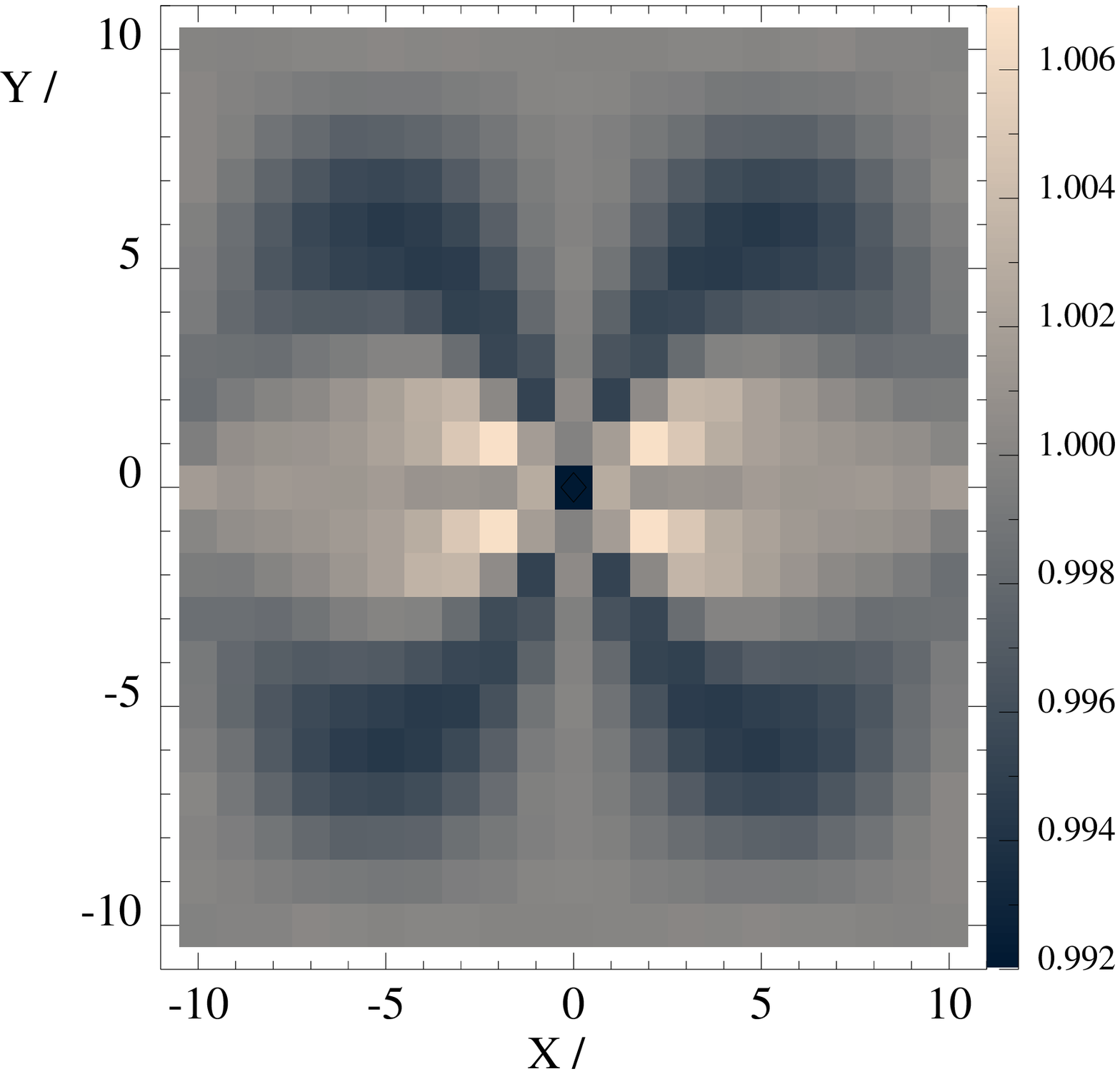}}
\subfigure[$u_x=0.2$]{\includegraphics[width=.3\textwidth]{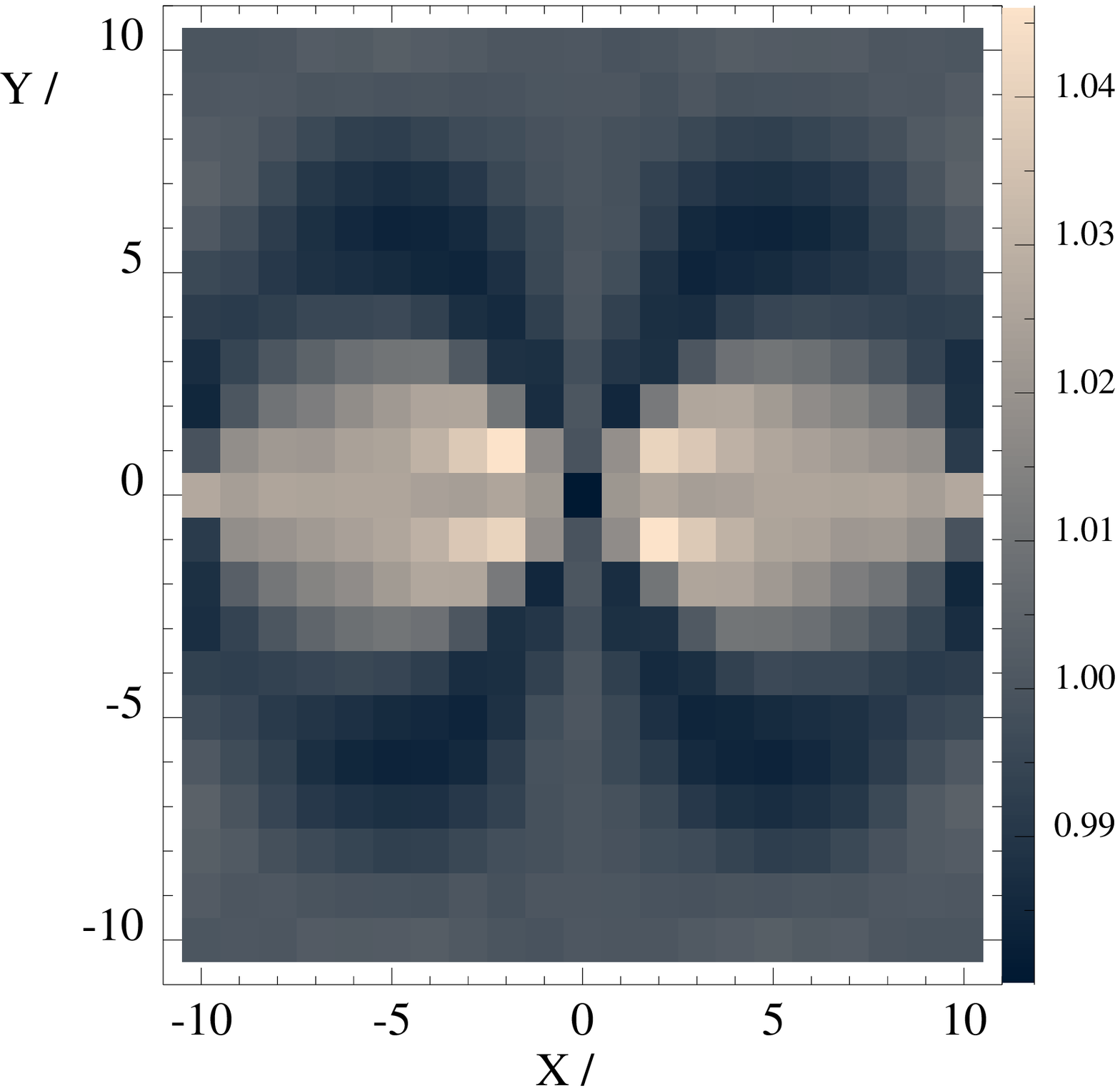}}
\end{psfrags}
\caption{Static structure factor $S_{\vecc{k}}(\tilde{j}_x)$ at different velocities measured for the $f$-norm with the look up table and $\Delta u_g = 0.02$.}
\label{fig:fsuxux}
\end{figure}

\begin{figure}[ht]
\begin{psfrags}
\psfrag{X /   }[Bl][Bc]{$k_x$}
\psfrag{Y /   }[Bl][Bc]{$k_y$}
%\subfigure[$u_x=0.0$]{\includegraphics[width=.3\textwidth]{fi_uxuy_u00.eps}}
%\subfigure[$u_x=0.1$]{\includegraphics[width=.3\textwidth]{wi_d2q9guuxuy.eps}}
\subfigure[$u_x=0.1$]{\includegraphics[width=.3\textwidth]{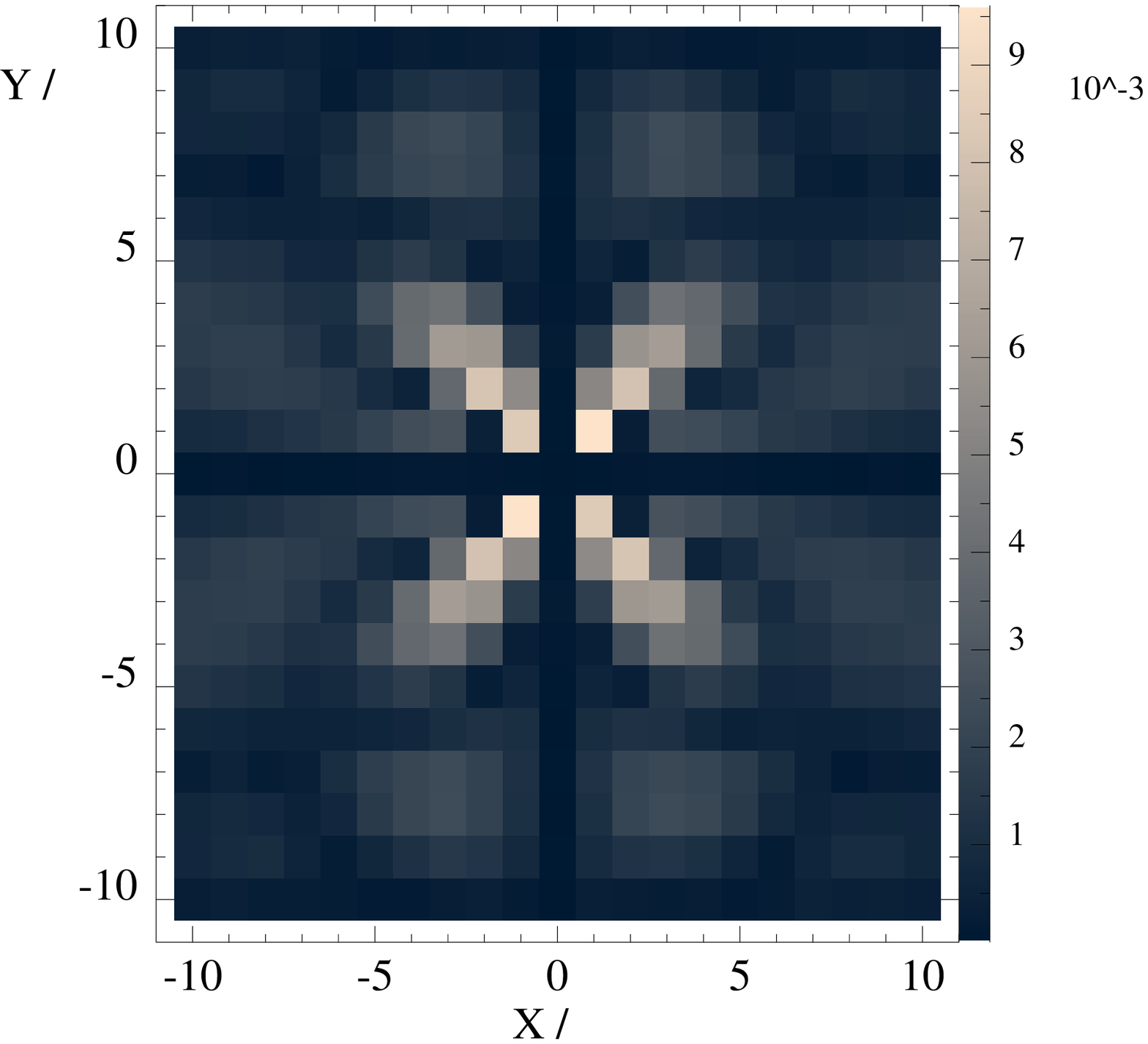}}
\subfigure[$u_x=0.2$]{\includegraphics[width=.3\textwidth]{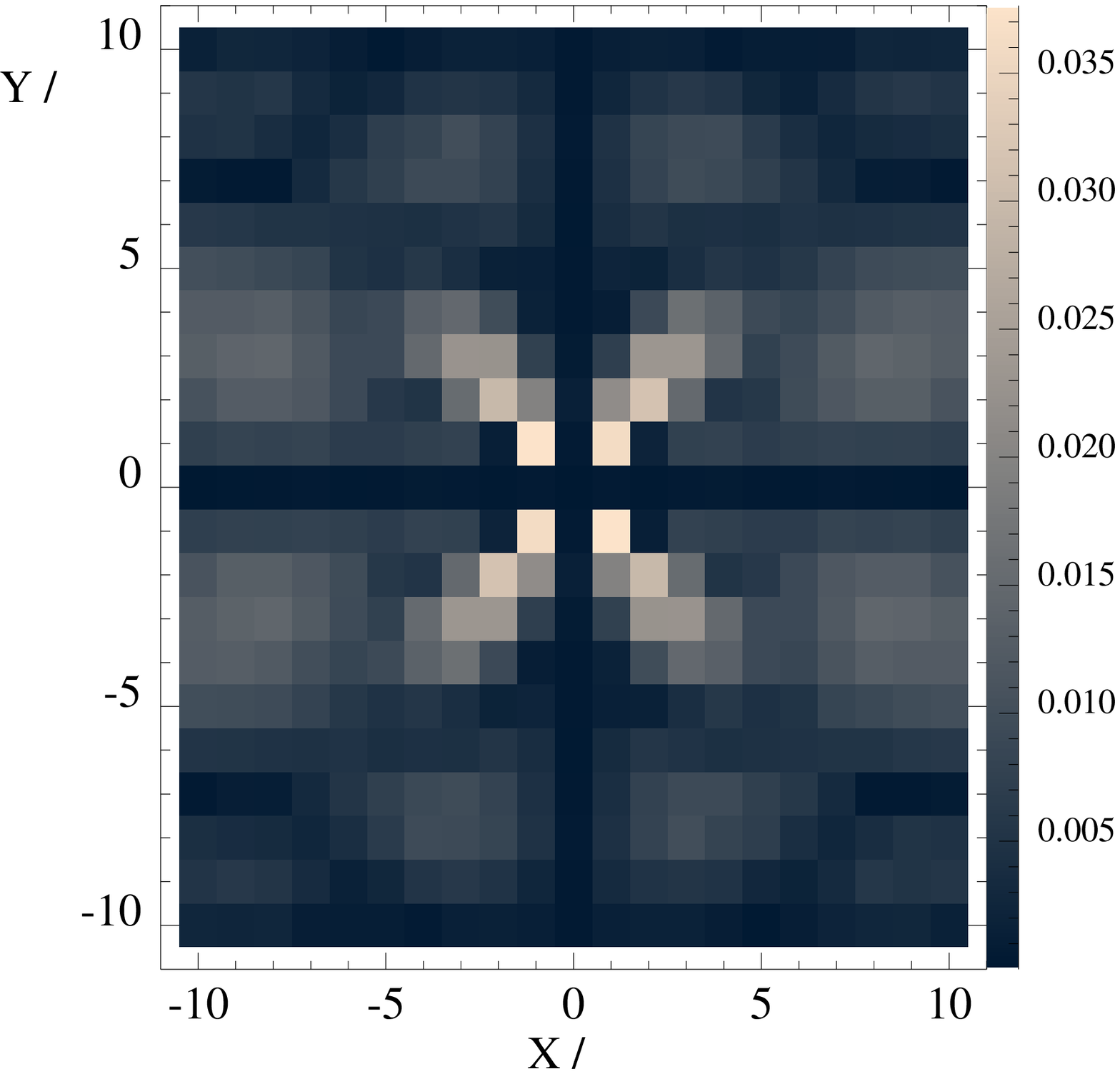}}
\end{psfrags}
\caption{Cross correlator $R_{\vecc{k}}(\tilde{j}_x, \tilde{j}_y)$ at different velocities measured for the $f$-norm with the look up table and $\Delta u_g = 0.02$.}
\label{fig:fsuxuy}
\end{figure}

Linear and quadratic fit coefficients for all moment correlators $\langle \delta \tilde{M}^a \delta \tilde{M}^b \rangle$ in Fig.~\figref{fig:fdmadmbfit1} show significant improvement as well. We notice that in particular the coefficients $l$ that apply to those off-diagonal correlators that have a linear dependence on $u_x$ at least a factor of~$13$ smaller than those measured in the Hermite norm case shown in Fig.~\figref{fig:wdmadmbfit1} (a). We also observe a decrease of the quadratic term $q$ but in line with the observations of Fig.~\figref{fig:fdmadma} the coefficients corresponding to some correlators decrease less compared to the ones observed in the Hermite norm in Fig.~\figref{fig:wdmadmbfit1} (b): $\langle  (\delta \tilde{\rho})^2\rangle $ from $1.9$ to $0.47$, $\langle (\delta \tilde{\Pi}_{xx-yy})^2 \rangle$ from $1.6$ to $0.54$, and $\langle (\delta \tilde{j}_y)^2 \rangle$ from $1.14$ to $0.75$.
% The $q$-coefficient of the cross correlator of $\langle \delta j_y \delta q_y\rangle$ actually increases from $0.20$ to $0.77$. It is the only one that increases.
%Overall we find that 

These findings are confirmed by the structure factor plots for the $f$-norm in Figs.~\figref{fig:fsrr}, \figref{fig:fsuxux}, and \figref{fig:fsuxuy} which for non-vanishing fixed velocity $u_{x,0}$ are significantly smaller than the one measured for the Hermite norm at the same velocity in Figs.~\figref{fig:wsrr},~\figref{fig:wsuxux}, and \figref{fig:wsuxuy}.

We can conclude that employing the $f$-norm significantly reduces the Galilean invariance effects observed on the Hermite norm implementation. The look up tables provide a practically feasible approach to implementing the $f$-norm at a performance loss of about 20 \%. All the measurements here were performed on a single CPU.

\section{Conclusion and Outlook}
The current standard implementation of thermal fluctuations in an isothermal ideal gas was tested for Galilean invariance violations. We found that with non zero average velocity the moment space covariance matrix of \eref{eqn:fluctf4} is neither diagonal nor are the diagonal elements unity as predicted and required by the derivation of the FDT in both \cite{adhikari-2005} and \cite{duenweg-2007}. We identified an approximation in the orthogonality condition that defines the moment space transforms \eref{eqn:fluctf3} as the likely source of the Galilean invariance violations as it directly removes an otherwise necessary velocity dependence from the moment space transforms. The approximation allows for the use of Hermite norm to define the moment space transforms. However, to recover Galilean invariance at least to some degree requires the matrix transforms to be locally velocity dependant, i.e. unique to every lattice site and the Hermite norm is no longer applicable. This led us to introduce a novel variant of the lattice Boltzmann method. We find that using the local fully velocity dependent $f$-norm to machine precision in a straight forward manner to be computationally impractical. Evaluating the individual matrix elements leads to an overhead in computational cost of $>2000\%$ in evaluating the individual matrix elements. However, as the Galilean invariance violations scale quadratically for most moments it is feasible to generate look up tables for the matrix elements on a velocity grid. This requires to projection of the equilibrium moments into the look up table reference velocity. This look up table approach provides comparable benefits to the locally orthogonalized transforms but at only a $20 \%$ loss of computation time. All the simulations presented here were performed in a example D$2$Q$9$ implementation. However, all calculations and considerations discussed can easily be generalized to other models. 
%The biggest practical difficulty would be the orthogonalization of the matrix elements for the $f$-norm for larger velocity sets. 
We provide a Mathematica notebook \cite{kaehler-2012-notebook} that contains the necessary calculations done for the D$2$Q$9$ model used here.
This new method is poentially important for non-equilibrium situations when locally varying flow fields exist which is the standard realm of lattice Boltzmann simulations.

\begin{acknowledgements}
The authors would like to thank Markus Gross and Eric Foard for helpful and insightful discussion.
This work has been funded, in part, by the ND EPSCoR SEED grand.
\end{acknowledgements}

%things needed:
%- lookup table
%- plots for lookup table method
%- moment definitions for lookup table method
%- comparison to w_i norm / quantification of improvement
%- equilibrium moments / projection into >eigenvelocity< representation

\appendix

%\section{d'Humieres' Multi-Relaxation transform for D2Q9}\label{sec:appdhumieres}
%\begin{widetext}

%One of the early MRT implementations was proposed by d'Humieres {\it et al.} \cite{dhumieres-1992}. Their D$2$Q$9$ transformation matrix is orthonormal in the sense of a Euclidean scalar product.

%\beq{eqn:dhumieresbase}
%\sum_i m_i^a f_i = 
%\left(
%\begin{array}{ccccccccc}
%\frac{1}{3} & \frac{1}{3} & \frac{1}{3} & \frac{1}{3} & \frac{1}{3} & \frac{1}{3} & \frac{1}{3} & \frac{1}{3} & \frac{1}{3}\\
%0 & \frac{1}{\sqrt{6}} & 0 & -\frac{1}{\sqrt{6}} & 0 & \frac{1}{\sqrt{6}} & -\frac{1}{\sqrt{6}} & -\frac{1}{\sqrt{6}} & \frac{1}{\sqrt{6}}\\
%0 & 0 & \frac{1}{\sqrt{6}} & 0 & -\frac{1}{\sqrt{6}} & \frac{1}{\sqrt{6}} & \frac{1}{\sqrt{6}} & -\frac{1}{\sqrt{6}} & -\frac{1}{\sqrt{6}}\\
%0 & \frac{1}{2} & -\frac{1}{2} & \frac{1}{2} & -\frac{1}{2} & 0 & 0 & 0 & 0\\
%0 & 0 & 0 & 0 & 0 & 1 & -1 & 1 & -1 \\
%-\frac{2}{3} &-\frac{1}{6} & -\frac{1}{6} & -\frac{1}{6} & -\frac{1}{6} & \frac{1}{3} & \frac{1}{3} &  \frac{1}{3} &  \frac{1}{3}\\
%0 &-\frac{1}{\sqrt{3}} & 0 & \frac{1}{\sqrt{3}} & 0 & \frac{1}{2\sqrt{3}} & -\frac{1}{2\sqrt{3}} & -\frac{1}{2\sqrt{3}} & \frac{1}{2\sqrt{3}}\\
%0 & 0 & -\frac{1}{\sqrt{3}} & 0 & \frac{1}{\sqrt{3}} & \frac{1}{2\sqrt{3}} & \frac{1}{2\sqrt{3}} & -\frac{1}{2\sqrt{3}} & -\frac{1}{2\sqrt{3}}\\
%\frac{2}{3} & \frac{1}{3} & \frac{1}{3} & \frac{1}{3} & \frac{1}{3} & \frac{1}{6} & \frac{1}{6} & \frac{1}{6} & \frac{1}{6}
%\end{array}
%\right) f_i = M^a
%\eeq

\section{Hermite norm D2Q9}\label{sec:apphermite}
For D2Q9 the equilibrium distribution employed is given by \eref{eqn:f0} with $\theta = 1/3$
\beq{eqn:d2q9f0} 
f_i^0(\rho, \vecc{u}, \theta) = \rho w_i \left\lbrack 1 + 3 \vecc{u}.v_i + \frac{9}{2}\left(\vecc{u}.v_i\right)^2 - \frac{3}{2}\vecc{u}.\vecc{u}\right\rbrack.
\eeq
The weights are given by 
\beq{eqn:d2q9w}
w_i = \left\{ 
\begin{array}{cl}
\frac{4}{9} & \text{if } i = 0\\
\frac{1}{9} & \text{if } i = 1, 2, 3, 4 \\
\frac{1}{36} & \text{if } i = 5, 6, 7, 8 
\end{array}
\right.
\eeq
In the case of the simple Hermite norm \eref{eqn:fluctf5} it is feasible to show the transformation matrices. The forward transform reads
\begin{widetext}
\begin{equation}
\label{eqn:d2q93a}
\sum_i m_i^a f_i = 
\left(
\begin{array}{ccccccccc}
 1 &  1 &  1 &  1 &  1 &  1 &  1 &  1 &  1 \\
 0 &  \sqrt{3} &  0 &  -\sqrt{3} &  0 &  \sqrt{3} &  -\sqrt{3} &  -\sqrt{3} &  \sqrt{3} \\
 0 &  0 &  \sqrt{3} &  0 &  -\sqrt{3} &  \sqrt{3} &  \sqrt{3} &  -\sqrt{3} &  -\sqrt{3} \\
 0 &  \frac{3}{2} &  \frac{-3}{2} &  \frac{3}{2} &  \frac{-3}{2} &  0 &  0 &  0 &  0 \\
 0 &  0 &  0 &  0 &  0 &  3 &  -3 &  3 &  -3 \\
 -1 &  \frac{1}{2} &  \frac{1}{2} &  \frac{1}{2} &  \frac{1}{2} &  2 &  2 &  2 &  2 \\
 0 &  -\sqrt{\frac{3}{2}} &  0 &  \sqrt{\frac{3}{2}} &  0 &  \sqrt{6} &  -\sqrt{6} &  -\sqrt{6} &  \sqrt{6} \\
 0 &  0 &  -\sqrt{\frac{3}{2}} &  0 &  \sqrt{\frac{3}{2}} &  \sqrt{6} &  \sqrt{6} &  -\sqrt{6} &  -\sqrt{6} \\
 \frac{1}{2} &  -1 &  -1 &  -1 &  -1 &  2 &  2 &  2 &  2
\end{array}
\right) f_i = M^a = \lpar
\begin{array}{c}
\rho \\
j_x\\
j_y\\
\Pi_{xx-yy}\\
\Pi_{xy}\\
\Pi_{xx+yy}\\
q_x\\
q_y\\
\epsilon
\end{array}
\rpar.
\end{equation}
%\end{widetext}

Likewise the back transform from moment space to velocity space is given by
%\begin{widetext}
\begin{equation}
\label{eqn:d2q93b}
\sum_a n_i^a M^a  = \left(
\begin{array}{ccccccccc}
 \frac{4}{9} &  0 &  0 &  0 &  0 &  \frac{-4}{9} &  0 &  0 &  \frac{2}{9} \\
 \frac{1}{9} &  \frac{1}{3\sqrt{3}} &  0 &  \frac{1}{6} &  0 &  \frac{1}{18} &  \frac{-1}{3\sqrt{6}} &  0 &  \frac{-1}{9} \\
 \frac{1}{9} &  0 &  \frac{1}{3\sqrt{3}} &  \frac{-1}{6} &  0 &  \frac{1}{18} &  0 &  \frac{-1}{3\sqrt{6}} &  \frac{-1}{9} \\
 \frac{1}{9} &  \frac{-1}{3\sqrt{3}} &  0 &  \frac{1}{6} &  0 &  \frac{1}{18} &  \frac{1}{3\sqrt{6}} &  0 &  \frac{-1}{9} \\
 \frac{1}{9} &  0 &  \frac{-1}{3\sqrt{3}} &  \frac{-1}{6} &  0 &  \frac{1}{18} &  0 &  \frac{1}{3\sqrt{6}} &  \frac{-1}{9} \\
 \frac{1}{36} &  \frac{1}{12\sqrt{3}} &  \frac{1}{12\sqrt{3}} &  0 &  \frac{1}{12} &  \frac{1}{18} &  \frac{1}{6\sqrt{6}} &  \frac{1}{6\sqrt{6}} &  \frac{1}{18} \\
 \frac{1}{36} &  \frac{-1}{12\sqrt{3}} &  \frac{1}{12\sqrt{3}} &  0 &  \frac{-1}{12} &  \frac{1}{18} &  \frac{-1}{6\sqrt{6}} &  \frac{1}{6\sqrt{6}} &  \frac{1}{18} \\
 \frac{1}{36} &  \frac{-1}{12\sqrt{3}} &  \frac{-1}{12\sqrt{3}} &  0 &  \frac{1}{12} &  \frac{1}{18} &  \frac{-1}{6\sqrt{6}} &  \frac{-1}{6\sqrt{6}} &  \frac{1}{18} \\
 \frac{1}{36} &  \frac{1}{12\sqrt{3}} &  \frac{-1}{12\sqrt{3}} &  0 &  \frac{-1}{12} &  \frac{1}{18} &  \frac{1}{6\sqrt{6}} &  \frac{-1}{6\sqrt{6}} &  \frac{1}{18} 
\end{array}
\right) M^a = f_i
\end{equation}
%\end{widetext}
where $n_i^a = m_i^a w_i$
The corresponding equilibrium moments $M^{a,0}$ are obtained directly by applying the forward transform to the equilibrium distribution. In the Hermite norm we find
\begin{equation}
\label{eqn:d2q94}
\begin{array}{ccccc}
\rho & = & M^{0,0} & = & \rho \\
j_x & = & M^{1,0} & = &  \sqrt{3}\rho u_x \\
j_y & = & M^{2,0} & = &  \sqrt{3}\rho u_y \\
\Pi_{xx-yy} & = & M^{3,0} & = &  \frac{3}{2} \rho (u_x^2 - u_y^2) \\
\Pi_{xy} & = & M^{4,0} & = &  3 \rho u_x u_y \\
\Pi_{xx+yy} & = & M^{5,0} & = &  \frac{3}{2} \rho (u_x^2 + u_y^2) \\
q_x & = & M^{6,0} & = &  0 \\
q_y & = & M^{7,0} & = &  0 \\
\epsilon & = & M^{8,0} & = &  0
\end{array}
\end{equation}.
\end{widetext}

\bibliography{../../literature/lbbib2}{}

\begin{thebibliography}{14}
\bibitem [{kae()}]{kaehler-2012-notebook}%
  \BibitemOpen
  \href@noop {} {}\bibinfo {note} {Mathematica notebook containing the full
  transforms, the full equilibrium moments and their third order expansion for
  D2Q9 is supplied as additional document with this publication. To evaluate
  the orthogonalizations a desktop computer with at least 4GB RAM is
  recommended.}\BibitemShut {Stop}%
\end{thebibliography}
\end{document}